\documentclass[10pt,twocolumn]{article}

\usepackage[utf8]{inputenc}
\usepackage{amsmath}
\usepackage{amsfonts}
\usepackage{amssymb}
\usepackage{graphicx}
\usepackage[obeyspaces,spaces]{url}
\usepackage{listings}
\usepackage{dsfont}

\ifx\titlerunning\undefined
\newcommand{\institute}[1]{#1}
\newcommand{\email}[1]{#1}
\fi

\newcommand{\inlineCode}{\texttt}
\newcommand{\figwidth}{0.4}
\newcommand{\programname}{Convino}
\newcommand{\Cpp}{C\kern-0.1em\raisebox{.2ex}+\kern-0.1em\raisebox{.2ex}+}

\newcommand{\PM}{pseu\-do-\-mea\-su\-re\-me\-nt}

\title{A method and tool for combining differential or inclusive measurements obtained with simultaneously constrained uncertainties}
\author{Jan Kieseler \institute{CERN, 1211 Geneva 23, \email{jan.kieseler@cern.ch}}}

\begin{document}

%\begin{frontmatter}

\ifx\titlerunning\undefined
\twocolumn[
\begin{@twocolumnfalse}
\maketitle%
\else
\titlerunning{Combination of measurements obtained with simultaneously constrained uncertainties} \authorrunning{Jan Kieseler}
\fi

\abstract{A method is discussed that allows combining sets of differential or inclusive measurements. 
It is assumed that at least one measurement was obtained with simultaneously fitting a set of nuisance parameters, representing sources of systematic uncertainties. As a result of beneficial constraints from the data all such fitted parameters are correlated among each other.
The best approach for a combination of these measurements would be the maximization of a combined likelihood, for which the full fit model of each measurement and the original data are required. 
However, only in rare cases this information is publicly available. 
In absence of this information most commonly used combination methods are not able to account for these correlations between uncertainties, which can lead to severe biases as shown in this article.
The method discussed here provides a solution for this problem. It relies on the public result and its covariance or Hessian, only, and is validated against the combined-likelihood approach. 
A dedicated software package implementing this method is also presented.
It provides a text-based user interface alongside a C++ interface. The latter also interfaces to ROOT classes for simple combination of binned measurements such as differential cross sections. 
}
\ifx\titlerunning\undefined
\vspace{1cm}
\end{@twocolumnfalse}
] 
\else
\maketitle
\fi

%\end{frontmatter}
\tableofcontents

\section{Introduction}

%Different measurements from experiments, blabla... 

%MISSING INFORMAITON OR NOT ACCESSIBLE FULL LIKELIHOOD. METHOD TO MAKE EQUIVALENT TO FULL LH COMBI

A common technique to reduce the impact of systematic uncertainties, in particular in precision measurements in high energy physics,  is to constrain the range of their variations by the data. A simultaneous fit of these variations and the parameters to be measured can be performed based on prior knowledge of the uncertainties and suitable distributions to constrain them. This technique can reduce the total uncertainty in many cases significantly, as in Refs.~\cite{Khachatryan:2137231,Sirunyan:2017uhy,Khachatryan:2016yzq}, and can be used to measure several parameters simultaneously (see e.g. Refs.~\cite{Kieseler:2015jzh,Aad:2014jra}).
However, it leads to non-negligible correlations between all fitted parameters. These are particularly important when at least one measurement that uses this technique contributes to a combination.

The most consistent approach to such combination would be to define a \textit{combined likelihood} based on the original models, including all systematic variations, and the original data the models were fit to. 
Also other commonly used combination techniques and the corresponding software tools would require this information~\cite{LYONS1988110,Nisius1,Nisius2,Alekhin:2014irh}, which is publicly available only in very rare cases and often unrecoverable. This poses a serious problem for a consistent combination involving measurements obtained with simultaneous nuisance parameter fits.

The method described here provides a solution for this problem, since it is based on the central results and their covariance or Hessians, only. 
It allows separating constraints and correlations imposed by the previously fitted data from those that stem from prior knowledge of the systematic variations.

Therefore, the combination can be performed accounting for correlations between the measurements as well as for correlations and constraints within each individual measurement. 

The dedicated software tool ``Convino'' is also presented in this article. It is specifically developed to perform combinations based on the method described here and provides a simple text-based user interface that can be used without knowledge of any programming language.
Assumptions on  correlations can be varied in an automated way.
Moreover, partially correlated measurements of different quantities (e.g. bins of a differential distribution) can be combined simultaneously accounting for all correlations.
In addition to the text-based interface, a \Cpp{} interface is provided to define the input to the combination. This interface can either read basic \Cpp{} standard library data types or  ROOT~\cite{Brun:1997pa} histogram and graph classes, which are commonly used in high-energy-physics analyses.

The combination method is described in Section~\ref{sec:math}.  It is validated against the combined likelihood approach in Section~\ref{sec:valid}. The effect of neglecting correlations within the same measurement is studied in Section~\ref{sec:negcorr}.
The installation and the user interface of the Convino program is described in Section~\ref{sec:install}. 

\section{Combination method}
\label{sec:math}

The combination is performed using a $\chi^2$ minimisation. The $\chi^2$ is defined as
\begin{equation}
\label{eq:fullchi}
\chi^2 =  \sum_{\alpha} \left(\chi^2_{s,\alpha} + \chi^2_{u,\alpha} \right) + \chi^2_p
\end{equation}
It is composed of three terms: the term $\chi^2_{s,\alpha}$ represents the results of each measurement $\alpha$ and its statistical uncertainties. It follows a Neyman or Pearson $\chi^2$ definition, with the statistical uncertainty being fixed for each measurement or being scaled with the combined value, respectively. A measurement can aim to determine a set of quantities, e.g. bins of a differential cross section, where the Pearson definition is more applicable. Alternatively a single quantity can be measured, e.g. the mass of a particle from a fit of an invariant mass peak position, where the Neyman definition is presumably better suited to describe the measurement. 
In both cases, the measured quantities are referred to as estimates in the following.
The additional term  $\chi^2_{u,\alpha}$ describes the correlations between the systematic uncertainties and constraints on them from the data for each measurement $\alpha$. The last term, $\chi^2_p$, incorporates prior knowledge of the systematic uncertainties and correlation assumptions between uncertainties of the measurements to be combined. 
%The individual terms are discussed in detail in the following subsections.

\ 

In a typical measurement that exploits simultaneous constraints on the uncertainties from the data, the sources of uncertainties are uncorrelated prior to the fit to the data and the knowledge about their variations is modelled by independent penalty terms in the original likelihood. 
The goal of the {me-thod} described here is to find an approximation for this likelihood that allows disentangling the independent penalty terms from the constraints and correlations between them, which are typically introduced by the data\footnote{The same procedure can be applied if correlations are present prior to the fit to the data}.
Therefore, the first central assumption is that the original likelihood for a measurement $\alpha$ can be approximated by:
\begin{equation}
\label{eq:evchi2}
\chi^2_\alpha  =  (\chi^2_{s,\alpha} + \chi^2_{u,\alpha}) + \sum_i (P_i^\alpha(\lambda_i) )^2 \text{,}
\end{equation}
where $(P_i^\alpha(\lambda_i) )^2$ represents the penalty term for a systematic uncertainty $i$ parametrised by a continuous parameter $\lambda_i$, such that $\lambda_i=1$ corresponds to a $1\sigma$ variation.
The terms $\chi^2_{s,\alpha}$ and $\chi^2_{u,\alpha}$ are defined as:
\begin{eqnarray}
\chi^2_{s,\alpha} & = &  \sum_{\mu\nu} {M}_{\mu\nu}^\alpha \frac{\xi_{\mu}^\alpha \xi_{\nu}^\alpha} { \tau_{\mu}^\alpha \tau_{\nu}^\alpha}  \text{ and} \\
\chi^2_{u,\alpha} & = &  \sum_{ij} \lambda_i  D_{ij}^\alpha \lambda_j   \text{, with} 
\end{eqnarray}
\begin{equation}
\xi_{\mu}^\alpha  =  x_{\mu}^\alpha -  \bar{X}_\mu \text{ and}
\end{equation}
\begin{equation}
\bar{X}_\mu  =   \bar{x}_\mu \prod_i (\lambda_i K_{\mu i}^\alpha /x_{\mu}^\alpha +1 ) + \sum_i \lambda_i  k_{\mu i}^\alpha
\text{.}
\end{equation}
Here, $x_{\mu}^\alpha$ is the estimate $\mu$ obtained in measurement $\alpha$ and $\bar{x}_\mu$ the combined value to be determined. The relation between both is given by $\tau_{\mu}^\alpha = ({\bar{X}_\mu/x_{\mu}^\alpha})^{0.5}$ for the Pearson $\chi^2$ definition. In case of the Neyman $\chi^2$, all $\tau_{\mu}^\alpha = 1$. The matrix $M$ represents the inverted statistical covariance of the estimates.
The parameters $k_{\mu i}^\alpha$ and $K_{\mu i}^\alpha$ model the effect of the systematic variations on the estimates for absolute or relative uncertainties, respectively. Here, a relative uncertainty is an uncertainty that scales with the measured value such as e.g. the luminosity uncertainty in a cross-section measurement, while absolute uncertainties have a constant value irrespective of the central result.
In principle, also other classes of dependencies can be incorporated through additional terms in $\xi^\alpha_\mu$.
The correlations between the uncertainties and the constraints that stem from the fit to the data are described by the matrix $D^\alpha$.  Throughout this paper, indices $\mu$ and $\nu$ are used for estimates, while systematic uncertainties are denoted with indices $i$ and $j$.

The procedure to obtain the parameters of $\chi^2_\alpha$ from the measurements to be combined are discussed in the following - firstly for results obtained through a simultaneous nuisance parameter fit and secondly for the specific case of orthogonal uncertainties.

\subsection{Measurements obtained with simultaneous fits}

The second central assumption of the method is that the parameters of $\chi^2_\alpha$ can be determined from the Hessian of measurement $\alpha$ evaluated at the best-fit values, $\tilde{H}^\alpha_\text{in}$.
The entries of the Hessian can be ordered such that the matrix can be split in the following sub-matrices:
\begin{equation}
\tilde{H}^\alpha_\text{in} = \begin{pmatrix}
\tilde{D} & {\kappa}^T \\
{\kappa} & \tilde{M}
\end{pmatrix}^\alpha \text{,}
\end{equation}
where $\tilde{M}$ describes the relation between the estimates $x_\mu^\alpha$ and $x_\nu^\alpha$. The relation between systematic variations and the estimates is described by $\kappa$. The matrix $\tilde{D}$ quantifies the relation between the systematic variations. 

All parameters of $ \chi^2_\alpha$ are determined by calculating analytically the Hessian of  $\chi^2_\alpha$, ${H}^\alpha (\vec{0})$, and identifying the resulting terms with their counterparts of the input $\tilde{H}^\alpha_\text{in}$. Here $\vec{0}$ means $\lambda_i =0$ and $x_\mu^\alpha - \bar{x}_\mu = 0\ \forall\ i,\ \mu$.
The components are calculated as follows:
\begin{equation}
 {H}^\alpha_{\mu\nu} (\vec{0}) = \frac{1}{2} \left(\frac{\partial^2 }{ \partial \Delta x_\mu^\alpha \partial \Delta x_\nu^\alpha}   \chi^2_\alpha \right)\biggr\rvert_{\vec{0}}
 =  {M}_{\mu\nu}  \text{,} 
\end{equation}
\begin{equation}
{H}^\alpha_{\mu i} (\vec{0}) = \frac{1}{2} \left( \frac{\partial^2 }{ \partial \Delta x_\mu^\alpha \partial \lambda_i}   \chi^2_\alpha \right)\biggr\rvert_{\vec{0}}
 =  \sum_\nu {M}_{\mu\nu} ( - \hat{k}_{\nu i}^\alpha) \text{,}
\end{equation}
with $\hat{k}_{\nu i}^\alpha = K_{\nu i}^\alpha + k_{\nu i}^\alpha$.
The matrix ${M}$ can be directly identified with $\tilde{M}$. 
Since $\tilde{M}$ stems from a measurement of a physics quantity, $\tilde{M}$ is positive definite and therefore invertible. Thus, the parameters $\hat{k}_{\nu i}^\alpha$ can be determined as:
\begin{equation}
\hat{k}_{\nu i}^\alpha = - \sum_\mu ((M^\alpha)^{-1})_{\mu\nu}\ \kappa_{\mu i}^\alpha \text{.}
\end{equation}
Since a variation $i$ is either relative or absolute, $\hat{k}_{\nu i}^\alpha$ equals either $K_{\nu i}^\alpha$ or $k_{\nu i}^\alpha$, with the other parameter being 0.
The terms describing the analytic relations between the systematic uncertainties are calculated as:
\begin{equation}
\begin{aligned}
{H}^\alpha_{i j} (\vec{0}) = & \frac{1}{2} \left( \frac{\partial^2 }{ \partial \lambda_i \partial \lambda_j}   \chi^2_\alpha \right)\biggr\rvert_{\vec{0}} \\
 =   &
D_{ij}^\alpha + \delta_{ij} \frac{1}{2} \frac{\partial^2 }{ \partial \lambda_i ^2}(P^\alpha_i)^2\biggr\rvert_{\lambda_i =0}  + \sum_{\mu \nu} M_{\mu \nu}^\alpha  \hat k_{\nu i}^\alpha\hat k_{\mu j}^\alpha \text{.} \\
\end{aligned}
\end{equation}
Only Gaussian penalty terms describing the prior knowledge of the uncertainties %or log-normal priors ($P_i(\lambda_i) = \ln (\lambda_i+1)$) 
are considered\footnote{In principle also other penalty terms (e.g. log-normal priors) can be accounted for. However, these lead to a non-constant $D^\alpha$.} ($P_i^\alpha(\lambda_i) = \lambda_i$), which simplify the equations, since in this case
\begin{equation}
\frac{1}{2}\frac{\partial^2 }{ \partial \lambda_i ^2}(P^\alpha_i)^2 \biggr\rvert_{\lambda_i =0} = 1 \text{.}
\end{equation}
In consequence $D^\alpha$ becomes:
\begin{equation}
D_{ij}^\alpha = \tilde{D}_{ij}^\alpha  -  \delta_{ij} -  \sum_{\mu \nu} M_{\mu \nu}^\alpha \hat k_{\nu i}^\alpha\hat k_{\mu j}^\alpha \text{,}
\end{equation}
such that all parameters of Eq.~\ref{eq:evchi2} are defined.

\subsection{Measurements with orthogonal uncertainties}
\label{sec:orthunc}

For orthogonal uncertainties, the same initial $\chi^2$ described in Eq.~\ref{eq:evchi2} is used. However, the calculation of its parameters does not necessarily require the full Hessian. Instead, the calculation of the parameters simplifies to:
\begin{equation}
\hat{k}_{\mu i}^\alpha = \frac{\sigma_{\mu i}^\alpha}{\sigma_{\mu \text{ total}}^\alpha} \text{,}
\end{equation}
with $\sigma_{\mu i}^\alpha$ being the contribution of uncertainty $i$ to the total uncertainty $\sigma_{\mu \text{ total}}^\alpha$ of estimate $x_\mu^\alpha$. The matrix $D^\alpha$ is 0, the terms of $M$ are calculated as:
\begin{equation}
M_{\mu \nu} = \frac{ \rho_{\mu\nu} }{\sigma_\mu \sigma_\nu} \text{.}
\end{equation}
Here, $\rho_{\mu\nu}$ is the statistical correlation between estimate $\mu$ and $\nu$, and $\sigma_\mu$ and $\sigma_\nu$  are the corresponding statistical uncertainties.

\ 

For the orthogonal uncertainties as well as for measurements obtained by simultaneous fits, the constraints from the prior knowledge of the uncertainties are implemented in Eq.~\ref{eq:fullchi} through the term: 
\begin{equation}
\label{eq:globalConstr}
\chi^2_{p} = \sum_{ij} P_i(\lambda_i) (C^{-1})_{ij} P_i(\lambda_j) \text{,}
\end{equation}
with $C$ being the matrix describing the correlation assumptions between the systematic uncertainties. In case no correlations are assumed, the term simplifies to:
\begin{equation}
\chi^2_{p}(\text{no corr}) = \sum_{i} P^2_i(\lambda_i)  \text{.}
\end{equation}
Only Gaussian penalty terms are considered in the following, such that $P_i(\lambda_i) = \lambda_i$.
For a combination, $C$ will be of the structure
\begin{equation}
C = \begin{pmatrix}
\mathds{1} & A & \cdots \\
A & \mathds{1} &  \cdots \\
\vdots & \vdots & \ddots \\
\end{pmatrix}\text{,}
\end{equation}
with matrices $A$ describing the correlation assumptions, and $\mathds{1} $ being the identity matrix.

\subsection{Technical implementation}
\label{sec:tech}

The final minimisation of Eq.~\ref{eq:fullchi} is performed using the Minuit algorithms~\cite{James:1975dr}. The total uncertainty on each combined value is determined by scanning $\chi^2 = \chi^2_\text{min} +1$ using the Minos algorithm. 
These algorithms as implemented in ROOT~6 as ``TMinuit2'' are employed.

The correlations that are assumed between systematic uncertainties can vary between -1 and 1. These extremes are special cases for which the correlation matrix $C$ becomes non-invertible. In practice, a correlation of $C_{ij}=\pm1$ means that  parameters $i$ and $j$ describe the same variation. %However, they are often defined as different parameters on input level and the correlation coefficient is assigned afterwards based on certain assumptions.
In such cases, an entry $C_{ij}=\pm1$ is replaced by $C_{ij}=\pm(1-10^{-3})$.
The difference to $\pm1$ is almost negligible. 

For illustration, $2\times 2$ parameters are chosen with $C_{ij}\approx \pm1$. The affected part of the $\chi^2$, $\chi^2_{F}$, can be simplified to
\begin{eqnarray}
\chi^2_{F} & = & \frac{1}{1-C_{ij}^2} (\lambda_i^2 + \lambda_j^2  \mp 2 C_{ij} \lambda_i \lambda_j)  \\
 & \approx & \frac{1}{1-C_{ij}^2} ( \lambda_i \mp \lambda_j)^2 
\end{eqnarray}
and corresponds to $( \lambda_i \mp \lambda_j)^2 \cdot 10^6$ for  $C_{ij} = \pm (1- 10^{-3})$. Given that a variation of $\lambda = \pm 1$ corresponds to only a fraction of the total uncertainty on each estimate, the effect of the approximation $C_{ij} = \pm (1- 10^{-3})$ is negligible. 

\section{Validation}
\label{sec:valid}

The validation is based on \textit{{\PM}s}. Each {\PM} is a binned likelihood fit with steerable central value and bin-wise uncertainties. 
This has the advantage, that the full likelihood of each {\PM} is known and can be adjusted to different scenarios. Therefore, it is possible to compare the results obtained with the method proposed here to the ones obtained using the combined likelihood as reference. Since the latter in principle contains arbitrarily more parameters, small deviations are expected.

The validation is first performed with respect to the statistical bias, only. Secondly, the modelling of systematic uncertainties is tested with respect to correlations between the uncertainties of the {\PM}s, and the modelling of relative uncertainties. 

%Each {\PM} is generated using a simultaneous binned Poisson-likelihood fit of the quantities to be determined ($%\bar{x}_\mu$) and randomly generated uncertainties with variations modelled by parameters $\lambda_i$.
For each {\PM} a Poisson likelihood is chosen to determine the central result $\bar{x}_\mu$. The bin-wise uncertainties are randomly generated and modelled by the parameters $\lambda_i$.
In the case that an uncertainty corresponds to an absolute variation, its effect on each bin is generated independently.
For more than one bin ($N_\text{bins}>1$) this results in correlations between the uncertainties after the fit, as well as in constraints on their variations. 

The likelihood for {\PM} $\alpha$ is defined as:
\begin{equation}
\label{eq:pseudoLH}
 L^\alpha = \prod_\mu  \prod_i^{N_\text{bins}^\alpha}  \mathcal{P}\left( X_\mu^\alpha, \bar{X}_{\mu i}^\alpha \right) 
 \cdot \prod_i \tilde{P}^\alpha_i(\lambda_i)  \text{,}
\end{equation}
with $\mathcal{P}$ being the Poisson likelihood and $\tilde{P}^\alpha_i(\lambda_i)$ the Gaussian penalty terms modelling the prior knowledge of each uncertainty. The parameters $X_\mu^\alpha$ and $\bar{X}^\alpha_{\mu i}$ are given as:
\begin{eqnarray}
X_\mu^\alpha & = &  \frac{x_\mu^\alpha}{N_\text{bins}^\alpha} \text{ and} \\
\bar{X}_{\mu i}^\alpha  & = &  \frac{\bar{x}_\nu}{N_\text{bins}^\alpha}\prod_j \left( \frac{K_{\nu j}^\alpha \lambda_j}{x_\nu}^\alpha + 1 \right) + \sum_j k_{\nu ij}^\alpha \lambda_j  \text{,} 
\end{eqnarray}
where $K_{\nu j}^\alpha$ describes the magnitude of global relative variations and $k_{\nu ij}^\alpha$ absolute shape variations, different for each bin $i$. The value of $x_\mu^\alpha$ is the input to each {\PM} and corresponds to the number of events that would be observed in a real measurement. The elements of the matrices $K^\alpha$ and $k_i^\alpha$ are chosen to describe different validation scenarios. Finally, the fit to determine $\bar{x}_\mu^\alpha$ is performed and the resulting Hessian is recorded.

The combined likelihood for several {\PM}s is given by:
\begin{equation}
L_\text{comb} = \left( \prod_\alpha \frac{L^\alpha}{\prod_i \tilde{P}_i^\alpha} \right) \cdot \phi (\lambda_0, ..., \lambda_N) \text{,}
\end{equation}
where $\phi$ models the prior knowledge of $N$ systematic uncertainties and the correlation assumptions between them,  analogue to $\chi^2_p$ in Eq.~\ref{eq:globalConstr}. For every validation step, the difference $\Delta \bar{x}$ between the result obtained with the method proposed in this document and using the combined likelihood  is recorded.
%This difference is normalised to the uncertainty on the combined value ($\Delta \bar{x}/ \sigma_{\bar{x}}$) to quantify the compatibility of both approaches.
The compatibility of both approaches is is quantified by $\Delta \bar{x}/ \sigma_{\bar{x}}$, which is the difference between their central results normalised to the total uncertainty of the combined-likelihood combination.

\subsection{Statistical bias}
\label{sec:statbias}

\begin{figure}[htbp]
  \begin{center}
\includegraphics[width=\figwidth\textwidth]{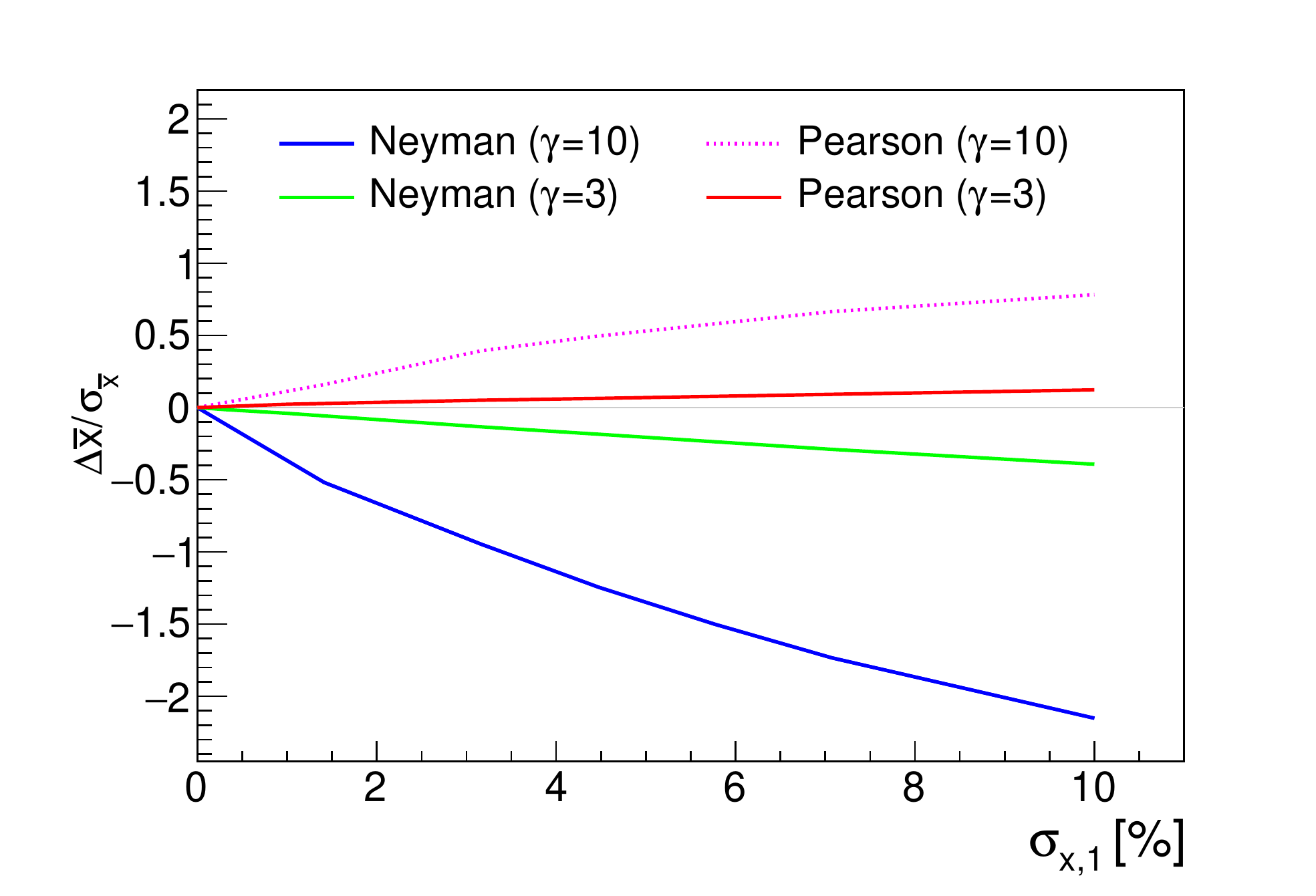}
\caption{Difference between the combined values using a direct Poisson-likelihood combination and the method proposed here with Neyman and Pearson $\chi^2$ definition relative to the total uncertainty. The estimates to be combined differ by about $\gamma\sigma$  and are displayed as a function of the first estimate's relative statistical uncertainty. The second estimate's statistical uncertainty scales accordingly. \label{fig:chibias}}
\end{center}
\end{figure}

To evaluate the statistical bias, the impact of systematic uncertainties on each {\PM} is set to 0, corresponding to $K=0$ and $k=0$. Only one quantity, $\bar{x}$, is determined from two estimates $x^a$ and $x^b$, chosen as:
\begin{eqnarray}
x^a & = & s \cdot 100 \text{,} \\
x^b & = & x^a + \gamma \sqrt{x^a} \text{,}
\end{eqnarray}
with $s$ being a scaling factor and $\gamma$ describing the compatibility between the estimates. The latter is chosen to be either $\gamma=10$ to describe two very incompatible measurements or $\gamma=3$ for a more realistic scenario where both estimates still agree well enough to enter a combination.
Two {\PM}s are generated for each choice of $s$ and $\gamma$ and are combined either using a Pearson or Neyman $\chi^2$ definition. 
For both choices, the uncertainties on the combined results agree very well with the ones obtained using the direct combination based on $L_\text{comb}$. The bias of the central value is shown in Figure~\ref{fig:chibias} relative to the uncertainty of the combined value. It behaves as expected: it is smaller but of opposite sign for the Pearson $\chi^2$ definition and is reduced with smaller statistical uncertainties and better compatibility between the results.

\subsection{Systematic uncertainties}
\label{sec:sysunc}

The effect of absolute systematic uncertainties is evaluated by combining two {\PM}s, with randomly chosen elements of the matrices  $k_\nu^\alpha$. An upper threshold $t$ is defined, such that for each element $i,\ j$:
\begin{equation}
| k_{\nu i j}^\alpha | \leq t \cdot X_{\mu}^\alpha \text{,}
\end{equation}
limiting the contribution of systematic uncertainties.
Two bins, two systematic uncertainties, and one $x_\mu^\alpha$ per {\PM} are considered. The sign of $k_{\nu i j}^\alpha$ is chosen to be constant for each systematic uncertainty. 
The estimates $x^a$ and $x^b$ for measurement $a$ and $b$ are set to:
\begin{eqnarray}
x^a & = & 30000 \text{ and}\\
x^b & = & 30600 %\text{.}
\end{eqnarray}
to reduce the effect of statistical uncertainties. The resulting statistical uncertainty of 0.6\% does not account for the difference of 2\% between both values, such that the modelling of the systematic uncertainties will affect the combination significantly.

For large systematic variations the maximisation of Eq.~\ref{eq:pseudoLH} with Minuit can become numerically unstable. This is the case when the variation becomes as large as the nominal entry, $X_\mu^\alpha$,  in at least one of the bins. Therefore, the Poisson likelihood is approximated with a Gaussian form, which is valid for low statistical uncertainties such as in this test. Thus, $L^\alpha$ becomes:\\
\begin{equation}
\label{eq:npseudoLH}
 L^\alpha = \prod_{\mu\nu}  \prod_i^{N_\text{bins}^\alpha} 
\exp\left[
-  S^\alpha_{\mu\nu} \frac{(\bar{X}_{\mu i}^\alpha - X_\mu^\alpha)(\bar{X}_{\nu i}^\alpha - X_\nu^\alpha)}{2 (\bar{X}_{\mu i}\bar{X}_{\nu i})^{1/2} }
\right] \text{.}
\end{equation}
The matrix $S^\alpha$ allows modelling direct statistical correlations between $\bar{X}_{\mu i}^\alpha$ and $\bar{X}_{\nu i}^\alpha$. Here, $S$ is set to $\mathds{1}$.

\begin{figure}[h]
  \begin{center}
\includegraphics[width=\figwidth\textwidth]{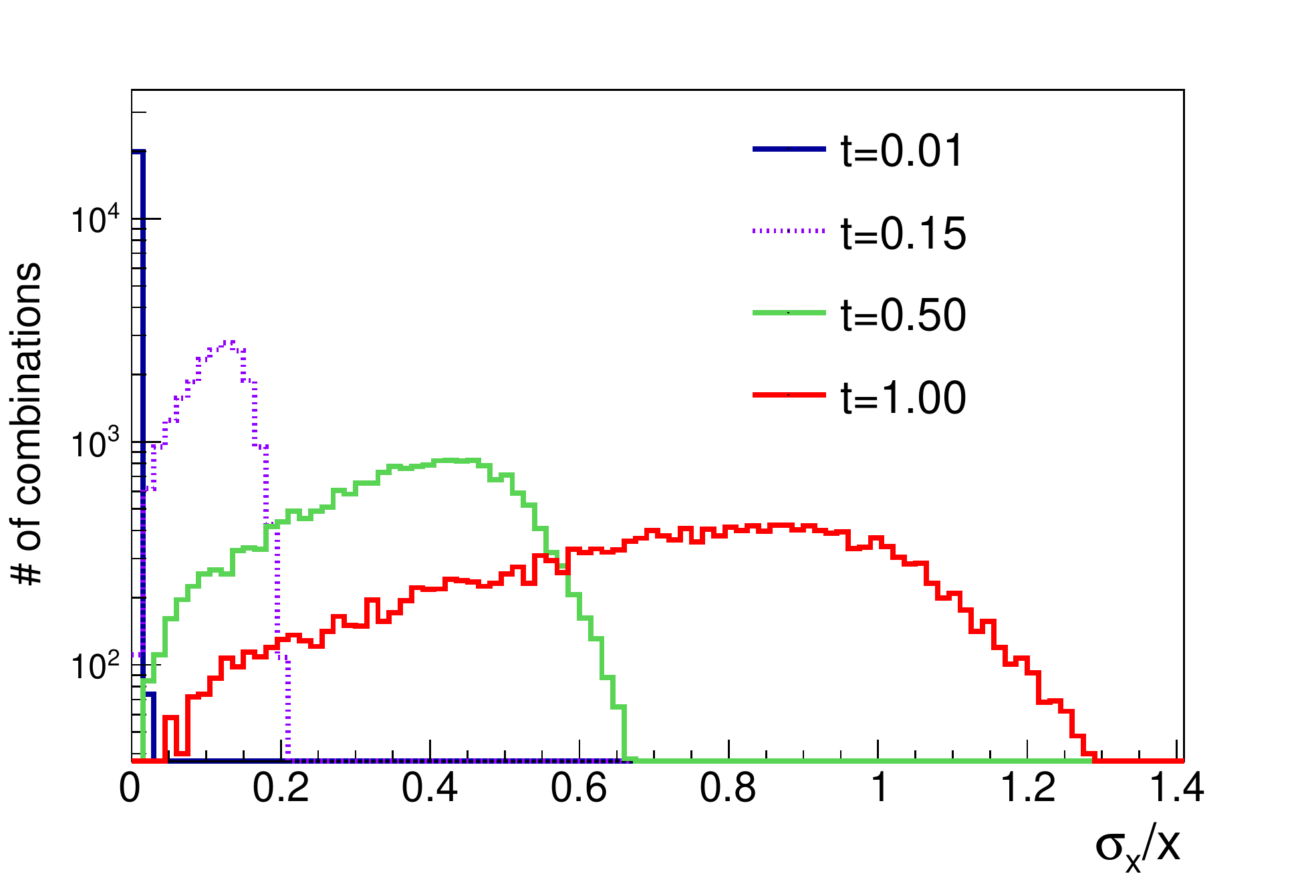}
\caption{Relative total uncertainty of the {\PM} $a$ for different values of the threshold $t$.  \label{fig:errb}}
\end{center}
\end{figure}

In total $2 \times 20,000$ {\PM}s are generated, each with a different random choice of the uncertainties.
The total relative uncertainty, $\sigma_x/x$, on the estimate of {\PM} $a$ is shown in Figure~\ref{fig:errb} for different values of the threshold $t$. Depending on $t$, the uncertainty varies from moderate values to more than 100\%. The same applies to {\PM} $b$ (not displayed). The average constraints on the systematic uncertainties reach from about 90\% ($t=0.01$) to 50\% ($t=1.0$) with respect to their initial $1\sigma$ variation.

\begin{figure}[h!]
  \begin{center}
\includegraphics[width=\figwidth\textwidth]{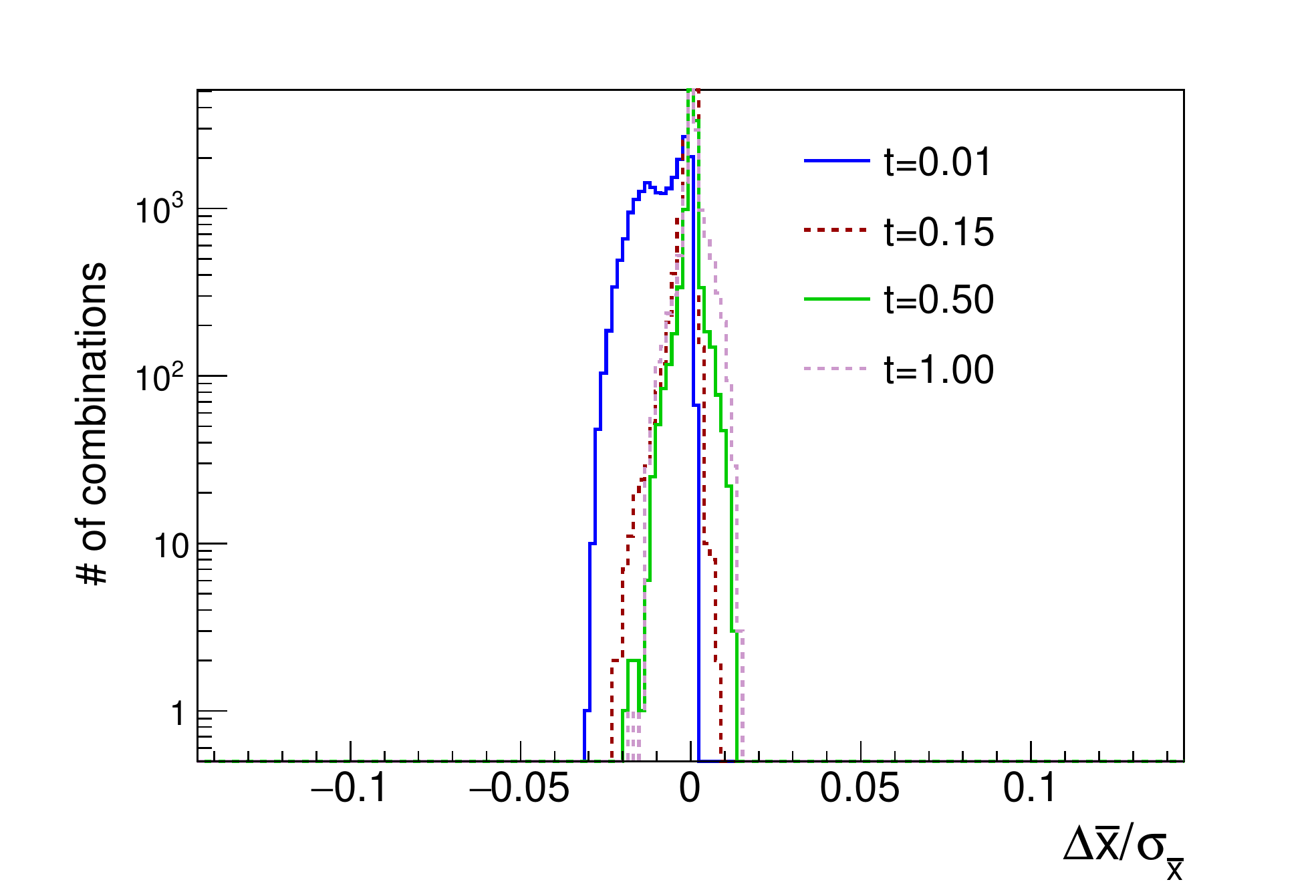}
\caption{
Difference between the combined result $\bar{x}$ obtained with the method proposed here and using a direct likelihood combination relative to the total uncertainty on $\bar{x}$, shown for different values of the upper threshold for systematic uncertainties $t$.\label{fig:deltaxbarrel}}
\end{center}
\end{figure}

In a first validation step, each uncertainty of one {\PM} is assumed to be highly correlated with exactly one uncertainty of the other {\PM} by assigning a correlation factor $c=0.99$. A total of 20,000 combinations are performed.
\begin{figure}[h]
  \begin{center}
\includegraphics[width=\figwidth\textwidth]{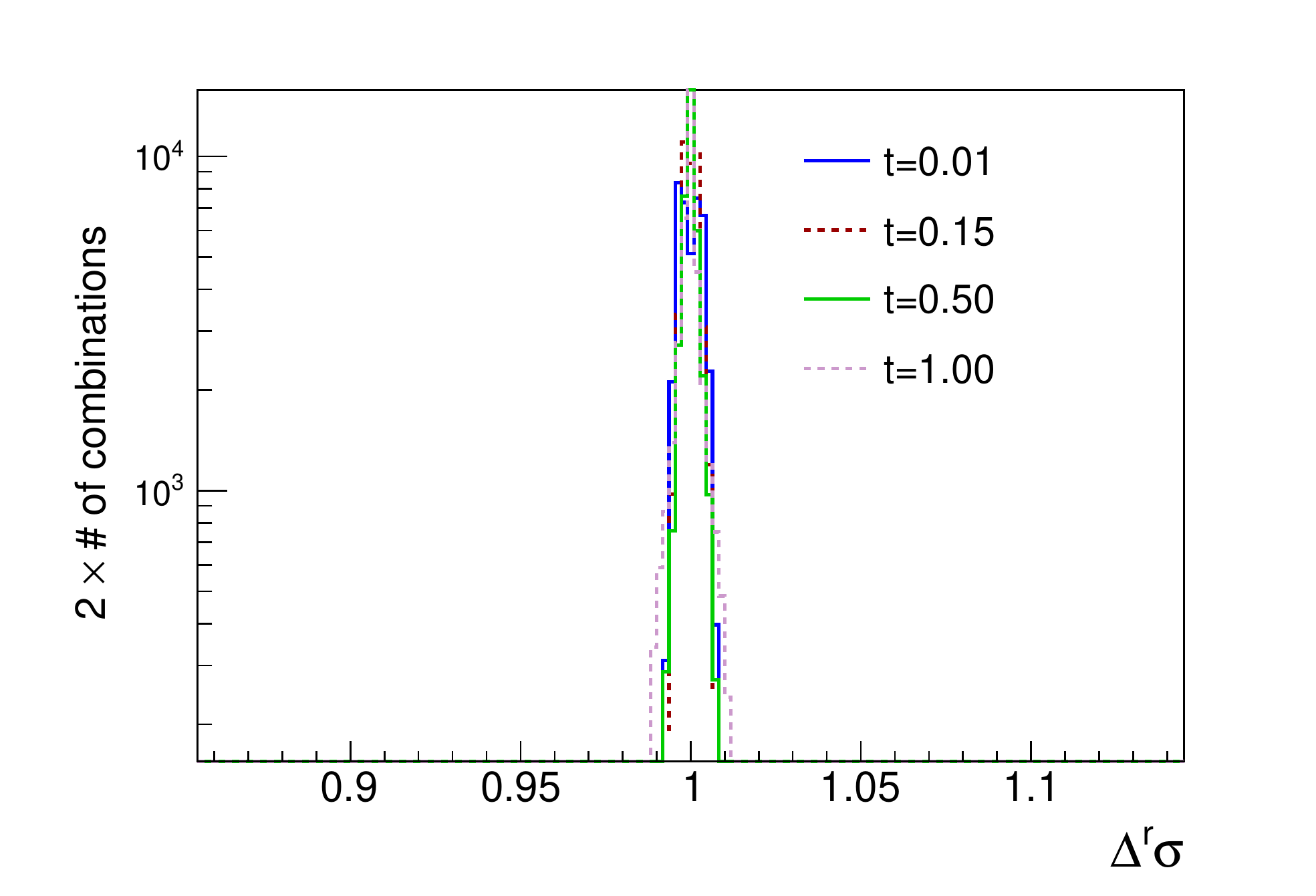}
\caption{Ratio of the uncertainties on the combined value obtained using the method proposed here and the direct likelihood combination, shown for different values of the upper threshold for systematic uncertainties $t$.  \label{fig:errb2}}
\end{center}
\end{figure}
The ratio $\Delta^r \sigma$ between the uncertainty on the combined value obtained with the method proposed here and by maximising $L_\text{comb}$ is shown in Figure~\ref{fig:errb2} as a function of $t$. Asymmetric uncertainties on the combined value are accounted for and are equally well described.
With an increasing contribution of the systematic uncertainties, the $\Delta^r \sigma$-distribution becomes slightly broader, but does not indicate any numerically relevant mismodelling. 
The resulting values for $\Delta \bar{x}/ \sigma_{\bar{x}}$ are illustrated in Figure~\ref{fig:deltaxbarrel}. 
%The distribution broadens slightly for larger values of $t$, but the effects are very small compared to the total uncertainty on the combined value.
For $t=0.01$, the statistical uncertainties are non negligible. This leads to a small bias towards lower values introduced by the choice of Eq.~\ref{eq:npseudoLH} and discussed in the Section~\ref{sec:statbias}. However, for all choices of $t$ and all pseudo experiments, the differences between the direct likelihood approach and the method described here are well below 5\% of the total uncertainty and can therefore be considered negligible.

\begin{figure}[h!]
  \begin{center}
\includegraphics[width=\figwidth\textwidth]{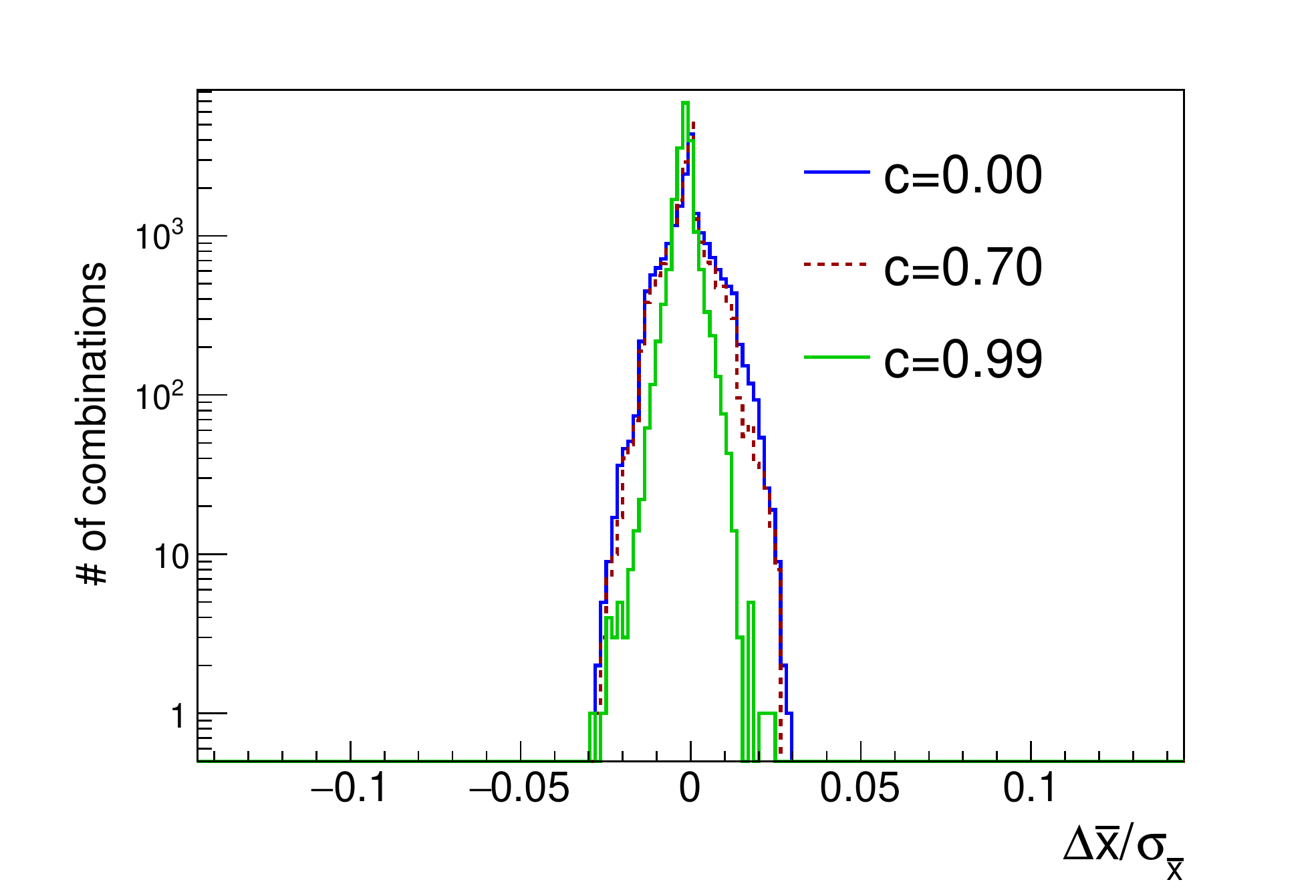}
\caption{
 Difference between the combined results $\bar{x}$ from the method proposed here and using a direct likelihood combination relative to the total uncertainty on $\bar{x}$. The distribution is shown for different values of the correlation $c$ between the systematic uncertainties of both {\PM}s.
\label{fig:syscorr}}
\end{center}
\end{figure}

\begin{figure}[htbp]
  \begin{center}
\includegraphics[width=\figwidth\textwidth]{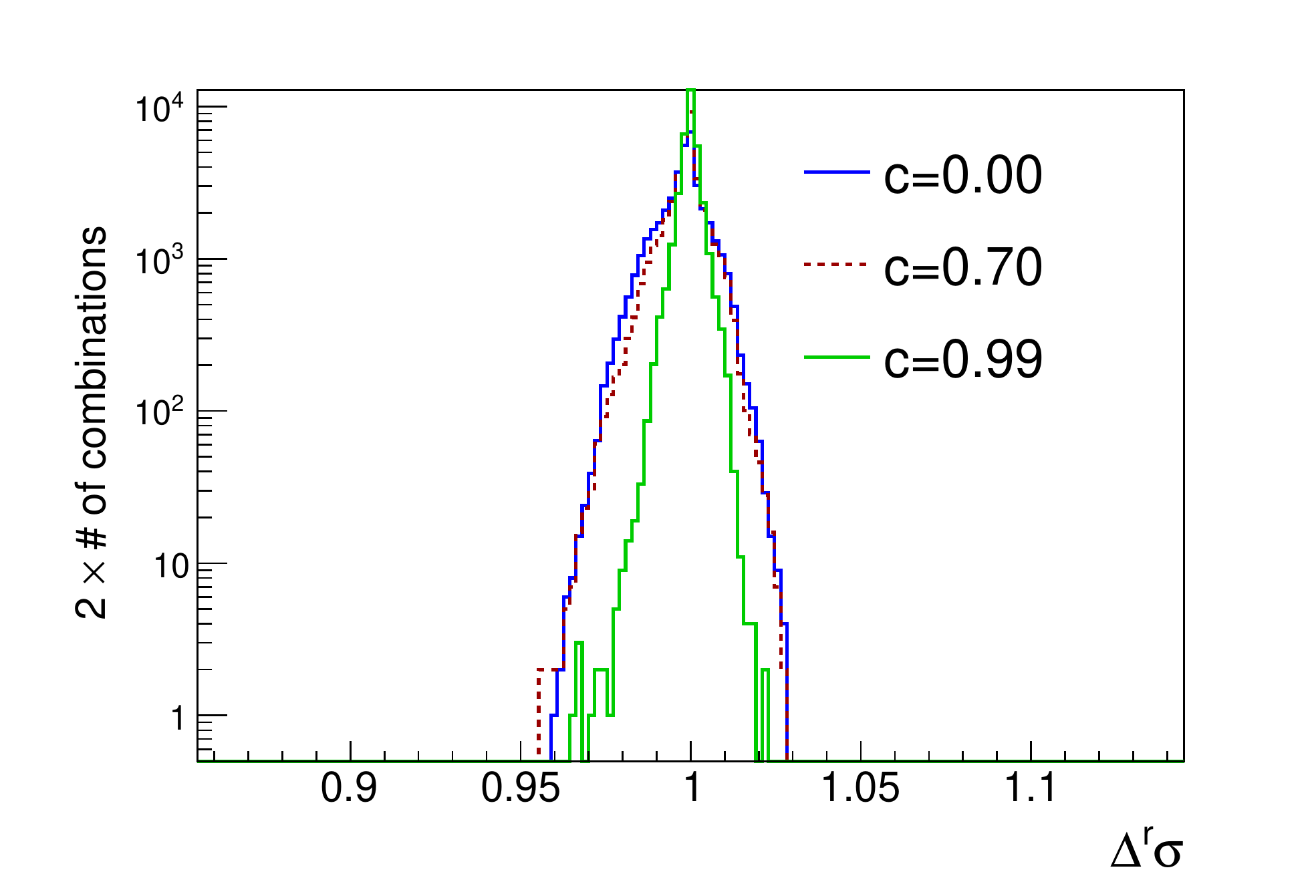}
\caption{
Ratio of the uncertainties on the combined value obtained using the method proposed here and the direct likelihood combination. The distribution is shown for different values of the correlation $c$ between the systematic uncertainties of both {\PM}s.
\label{fig:syscorr2}}
\end{center}
\end{figure}

Moreover, the dependence on the assumed correlation between the uncertainties of both {\PM}s is studied, as well as possible biases with respect to the number of bins in each {\PM}. 
Figure~\ref{fig:syscorr} shows the dependence of $\Delta \bar{x}/ \sigma_{\bar{x}}$ on the choice for the correlation coefficients $c$ for $t=1$. 
The modelling worsens slightly when $|c|$ decreases, but is below about 3\% with respect to the total uncertainty on the combined value for all 20,000 pseudo experiments. 
Also, the total uncertainty remains well modeled with only a very moderate increase of combinations with $|\Delta^r \sigma |$ slightly different from 1, as shown in Figure~\ref{fig:syscorr2}. The same conclusion can be drawn when the procedure described here is repeated for a different number of bins in each {\PM} (not shown here). All results for 2, 4, 20, and 100 bins show a good modelling of the combined likelihood approach with respect to the central values and the total uncertainties.

In general, the central result and its uncertainty are very well modelled for a large range of relative contributions from systematic uncertainties, correlations among them, and the chosen number of bins in each {\PM}. 
Very small deviations of the order of a few per-cent with respect to the total uncertainty are observed.
These are expected as a result of reducing the binned information of the initial measurement likelihood to one or many estimates and a corresponding Hessian.
The method described here shows a similar stability for different choices of $x^a$ and $x^b$, and of the number of uncertainties. Moreover, it is also valid for multiple estimates within one {\PM} without statistical correlations between them. The case of statistical correlations is discussed separately in Section~\ref{sec:dircorr}.

\subsection{Modeling of statistical correlations}
\label{sec:dircorr}

\begin{figure}[htbp]
  \begin{center}
\includegraphics[width=\figwidth\textwidth]{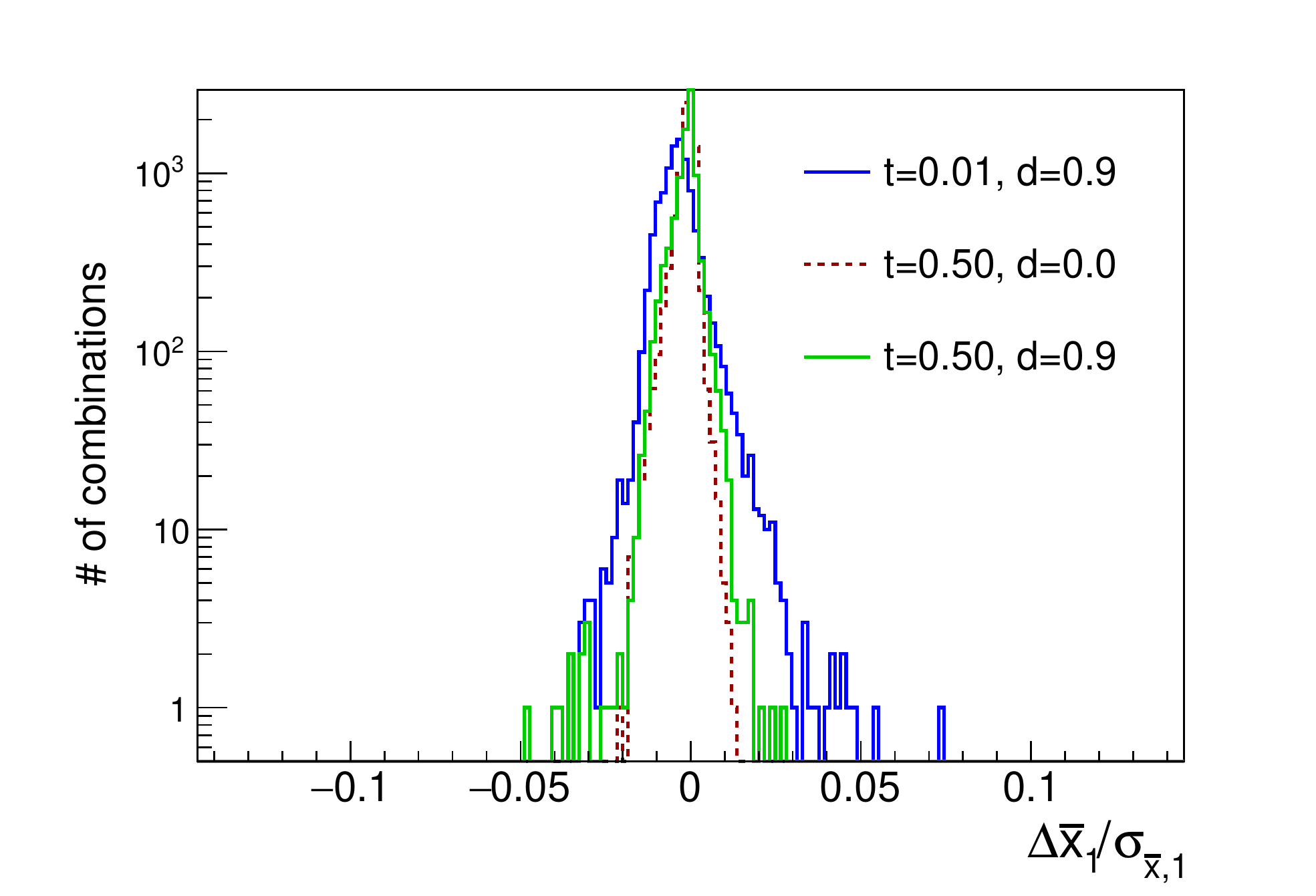}
\caption{
Difference between the combined value of $\bar{x}_1$ obtained with the method described here and using a direct likelihood combination relative to the total uncertainty of $\bar{x}_1$. The distribution is shown for different values of the scale $t$ for systematic uncertainties and the statistical correlation between the estimates $d$. 
The combination assumes no correlation between the uncertainties of both {\PM}s.
 \label{fig:dircorr}}
\end{center}
\end{figure}

\begin{figure}[htbp]
  \begin{center}
\includegraphics[width=\figwidth\textwidth]{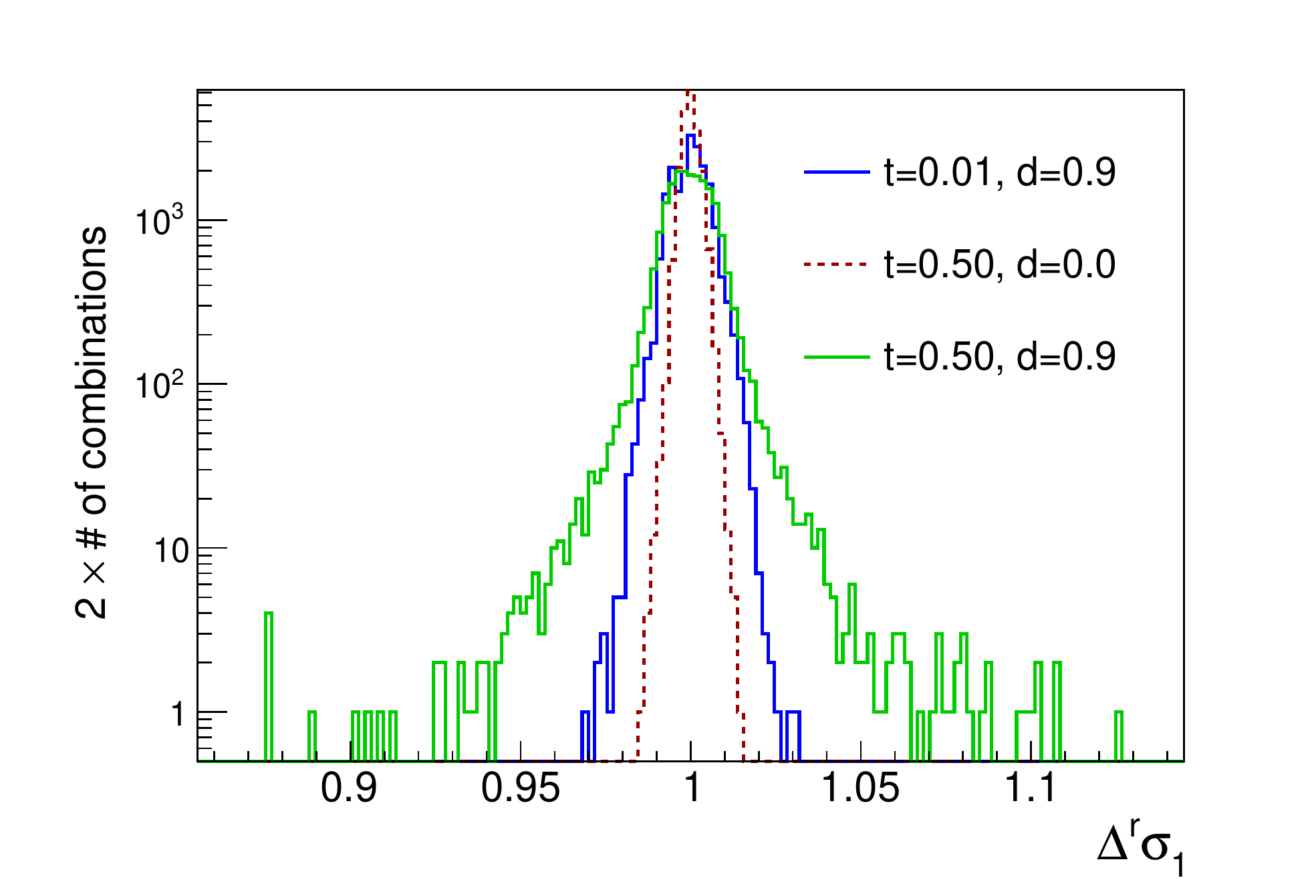}
\caption{
Ratio of the uncertainties on $\bar{x}_1$ obtained with the method proposed in this document and using a direct likelihood combination.  The distribution is shown for different values of the scale for systematic uncertainties $t$ and the statistical correlation between the estimates $d$. The combination assumes no correlation between the uncertainties of both {\PM}s.
 \label{fig:dircorr1a}}
\end{center}
\end{figure}

The correct modelling of statistical correlations between the estimates within a measurement is tested by generating two {\PM}s $a$ and $b$ similar to Section~\ref{sec:sysunc}, each with two estimates $x^a_1$ and $x^a_2$ or $x^b_1$ and $x^b_2$, respectively. The corresponding correlation matrices $S^a$ and $S^b$ are randomly chosen to have off-diagonal elements with an absolute value of $d\pm 0.1$. In total, 10000 combinations are performed for each choice of $d=\{0,\ 0.3,\ 0.9\}$, $t=\{0.01,\ 0.50\}$, and $c=\{0.00, 0.99\}$. The values for $x^\alpha_\mu$ are chosen to be $x^a_1  =  30000 $, $x^b_1 =  30600$, $x^a_2  =  20000$, and $x^b_2  =  20500$.
The resulting values for $\Delta \bar{x}_1/ \sigma_{\bar{x},1}$ and $\Delta^r\sigma_1$ are illustrated in Figure~\ref{fig:dircorr} and~\ref{fig:dircorr1a} for $c=0$. 

No significant mismodelling of the statistical correlation between estimates of the same measurement can be observed. 
The dependence on  $c$ is similar to the one discussed in Section~\ref{sec:sysunc} and not shown here. Also the result of the combination of $x_2$ shows identical behaviour and is therefore not depicted either.  Different choices for $x^\alpha_\mu$ were tested and confirm a good modelling with respect to $d$.

\subsection{Relative uncertainties}

\begin{figure}[h]
  \begin{center}
\includegraphics[width=\figwidth\textwidth]{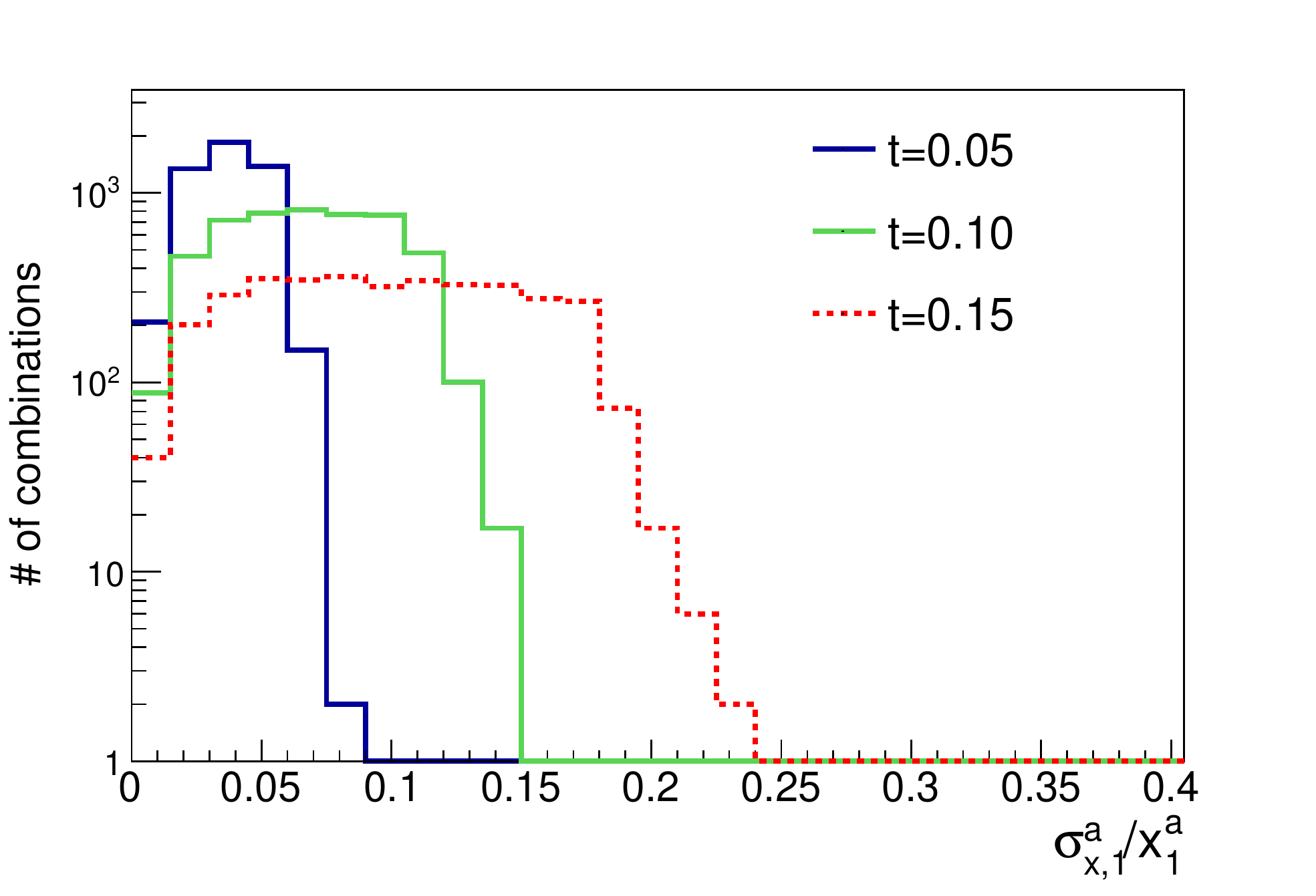}
\caption{Relative total uncertainty of the {\PM} $a$ for different values of the threshold $t$ for systematic uncertainties.  All {\PM}s comprise one relative and two absolute uncertainties.
 \label{fig:relerrrelunc}}
\end{center}
\end{figure}

\begin{figure}[h]
  \begin{center} 
\includegraphics[width=\figwidth\textwidth]{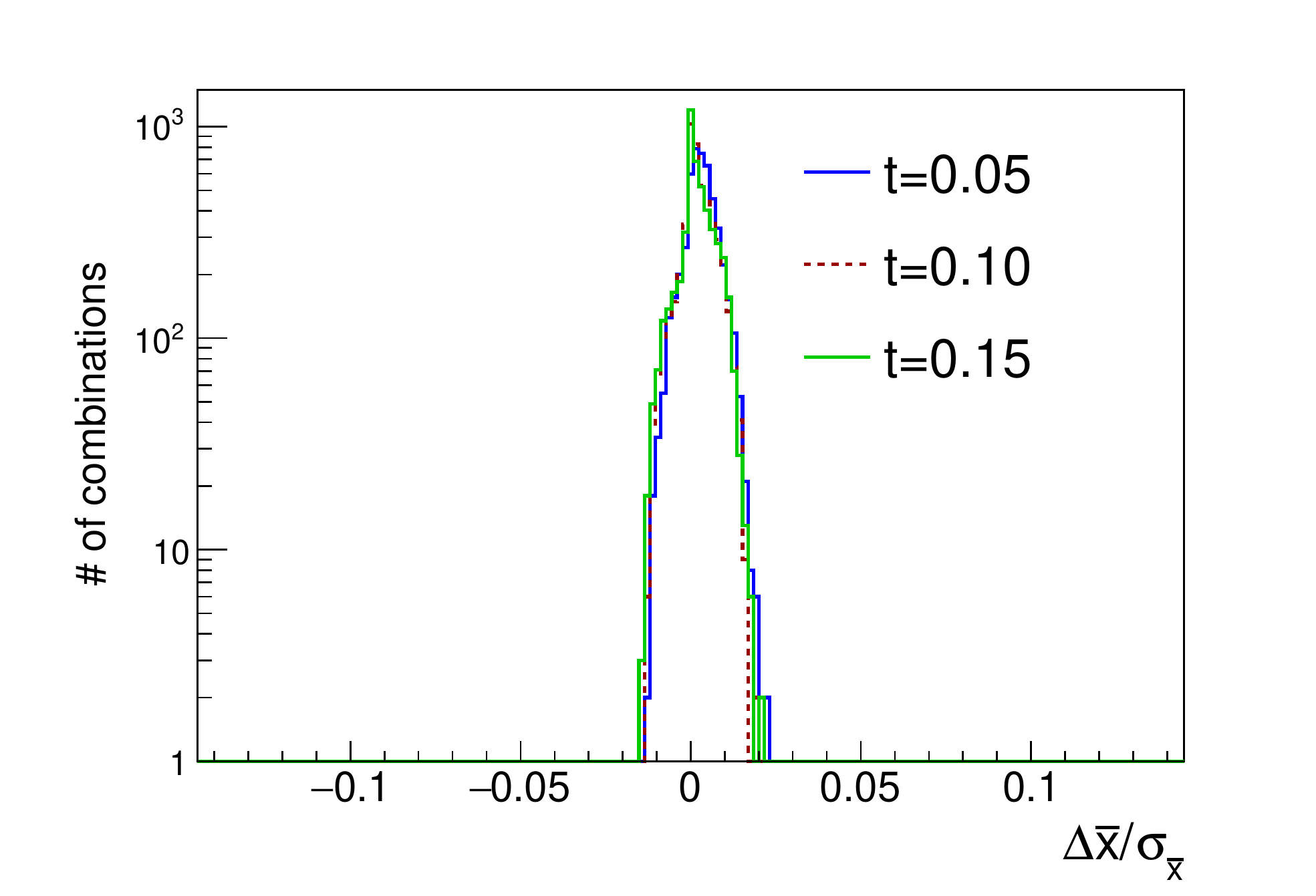}
\caption{Difference between the combined value of $\bar{x}$ obtained with the method proposed here and using a direct likelihood combination relative to the total uncertainty of $\bar{x}$, shown for different values of the threshold $t$ for systematic uncertainties.  All {\PM}s comprise one relative and two absolute uncertainties.
 \label{fig:relerrrelunc2}}
\end{center}
\end{figure}

\begin{figure}[h]
  \begin{center}
\includegraphics[width=\figwidth\textwidth]{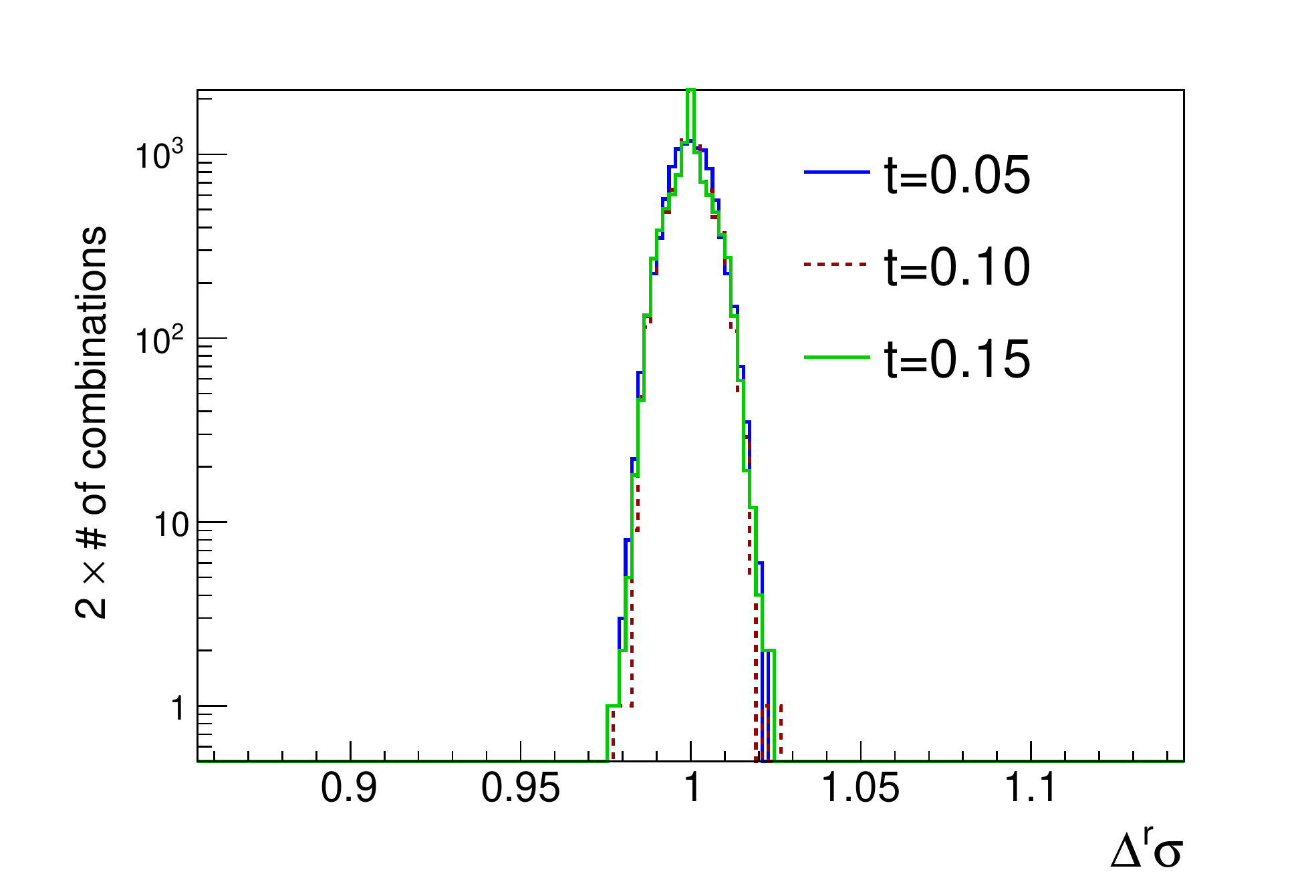}
\caption{Ratio of the uncertainties on $\bar{x}$ obtained with the method described in this document and using a direct likelihood combination, shown for different values of the threshold $t$ for systematic uncertainties.  All {\PM}s comprise one relative and two absolute uncertainties.
 \label{fig:relerrrelunc3}}
\end{center}
\end{figure}

The modelling of relative uncertainties is studied by generating two {\PM}s, each of them with one parameter to be combined, one relative uncertainty, and two absolute uncertainties. The relative uncertainty applies to all bins in the same way and will therefore not receive constraints. In consequence, it will be dominant. Thus, the total uncertainty of each {\PM} will differ from the dependence on $t$ previously illustrated in Figure~\ref{fig:deltaxbarrel}. A total of $2 \times 5000$ {\PM}s are generated.
Figure~\ref{fig:relerrrelunc} shows the relative uncertainty of {\PM} $a$, including one relative uncertainty, as a function of $t$. For $t$ larger than 0.15, the direct likelihood combination shows instabilities in some cases, likely related to the Gaussian penalty terms, while log-normal terms would be more suitable for large relative uncertainties. 
As shown in Figures~\ref{fig:relerrrelunc2} and~\ref{fig:relerrrelunc3}, also when combining {\PM}s with contributions from relative uncertainties, central values and uncertainties are well modelled, assuming the uncertainties of one {\PM} to be uncorrelated with the uncertainties of the other. The same holds true for high correlations between the {\PM}s. 

Additionally, the method is validated using exactly one estimate per {\PM} and one large relative uncertainty of +15\%. The input estimates are set to $30000 + \beta$, where $\beta$ is a randomly generated value between 0 and 750.
The uncertainty is assumed to be fully correlated between the {\PM}s. This results in asymmetric uncertainties on each {\PM} and the combined value. Moreover, for this particular choice of uncertainties, the combined value can be larger than the highest input estimate.
When comparing the direct likelihood combination to the method proposed here, also in this case no bias with respect to the central value or its uncertainties can be observed.

In summary, the combination method described here does not require the original data and the full fit model, but sufficiently describes the initial measurement for a large variety of possible central values, binning choices and uncertainties. In consequence, the combination results are numerically equivalent to a using the full likelihood information, in particular in case of dominant systematic uncertainties.

\ 

\section{Neglecting correlations}
\label{sec:negcorr}

\begin{figure}[hbtp]
  \begin{center}
\includegraphics[width=\figwidth\textwidth]{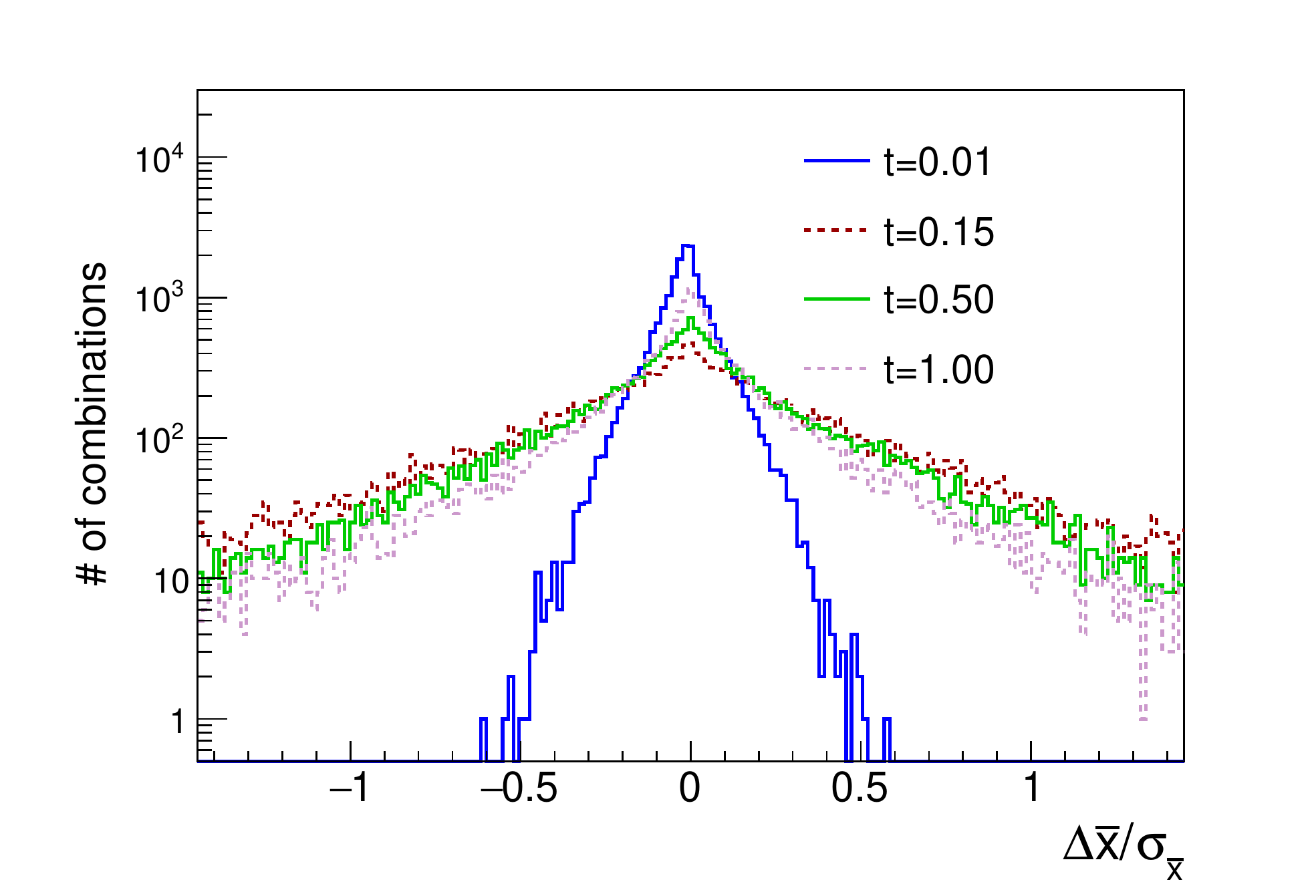}
\caption{
 Difference between the combined results neglecting correlations between uncertainties within a measurement and using a direct likelihood combination relative to the total uncertainty on $\bar{x}$.
The distribution is shown for different values of the upper threshold for systematic uncertainties $t$. For comparison with the method proposed here, see Figs.~\ref{fig:errb2} and~\ref{fig:deltaxbarrel}.
\label{fig:nocorr}}
\end{center}
\end{figure}

\begin{figure}[hbtp]
  \begin{center}
\includegraphics[width=\figwidth\textwidth]{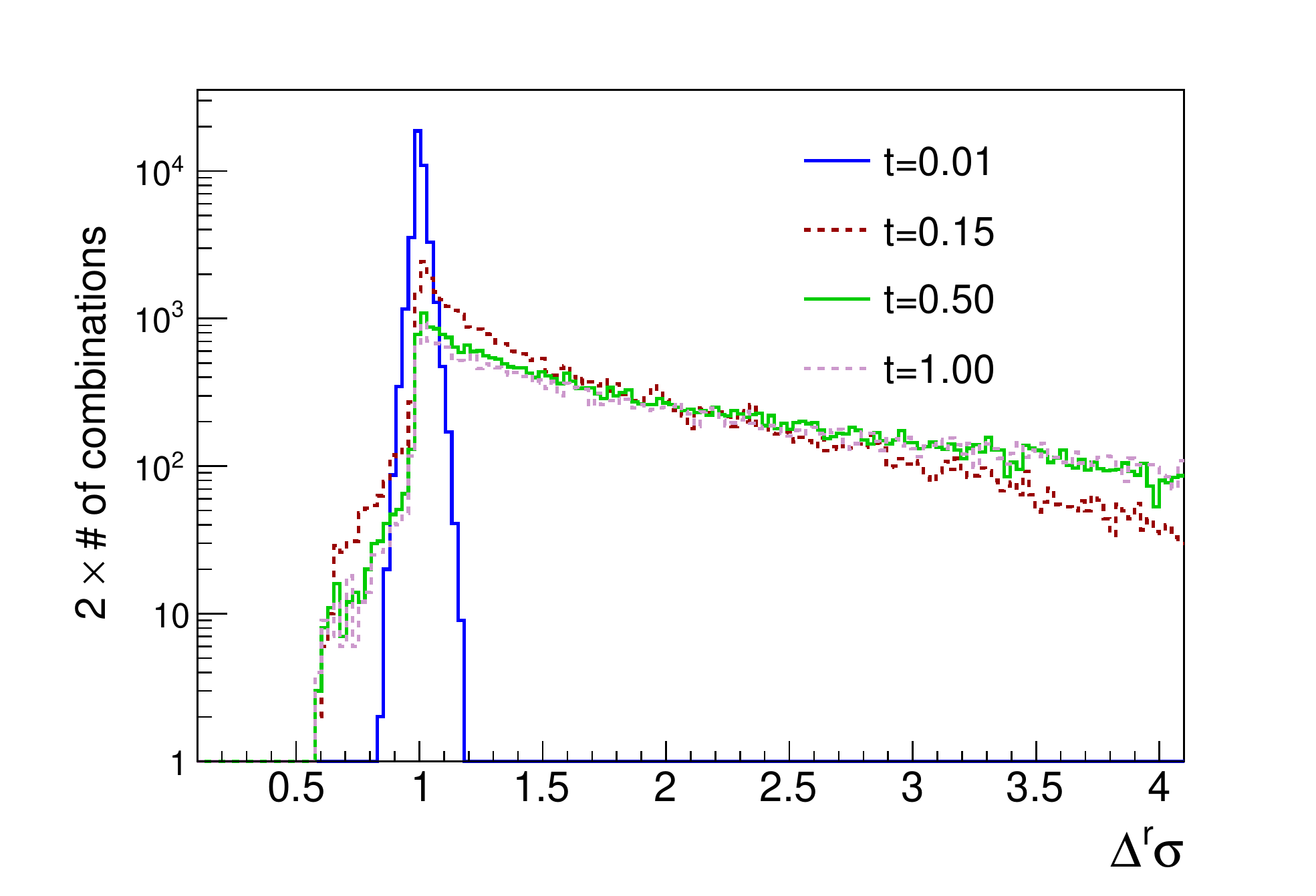}
\caption{
 Ratio of the uncertainties on the combined value obtained neglecting correlations between uncertainties within a measurement and using a direct likelihood combination. The distribution is shown for different values of the upper threshold for systematic uncertainties $t$. For comparison with the method proposed here, see Figs.~\ref{fig:errb2} and~\ref{fig:deltaxbarrel}.
\label{fig:nocorr2}}
\end{center}
\end{figure}

For comparison, two {\PM}s $a$ and $b$ with two bins, two systematic uncertainties, and the same parameters described in Section~\ref{sec:sysunc} are combined neglecting correlations between uncertainties within the same {\PM}, but still considering strong correlations between {\PM}s $a$ and $b$. This approximates the situation in which the BLUE method~\cite{LYONS1988110,Nisius1,Nisius2} can be used  for the combination.
The correlations within one {\PM} are removed by inverting $(D^\alpha + \mathds{1})$ in Eq.~\ref{eq:evchi2}, removing the off-diagonal elements of the resulting covariance matrix, and replacing $D^\alpha$ by the inverse of this covariance matrix minus $\mathds{1}$. By choosing the Neyman $\chi^2$ definition in addition, this makes this test equivalent to the BLUE method.
As shown in Figure~\ref{fig:nocorr}, this approximation can lead to wrong individual combination results with respect to the central value when the contribution of systematic uncertainties becomes non-negligible. 
Also, the uncertainty on the combined value can be severely mismodelled if the correlations within one measurement are neglected, as displayed in Figure~\ref{fig:nocorr2}. The total uncertainty can be underestimated or strongly overestimated, in particular if it is dominated by systematic uncertainties.
Therefore, it is crucial to model these correlations consistently when performing a combination of results obtained in simultaneous fits of systematic uncertainties and the quantity to be determined.

\

\section{Program Installation and User Interface}
\label{sec:install}

The method described in Section~\ref{sec:math} is implemented in the dedicated Convino program for the combination of experimental results.
The source code can be found at \path{https://github.com/jkiesele/Convino/releases}.  It can be compiled using \inlineCode{make} with gcc version~$4.9$ or newer, or clang~8.0.0 or newer (OSX) and ROOT~6 installed on the system. Other versions might be sufficient but are not tested.

The measurements and the configuration for the combination are contained in human-readable text files.
Alternatively, a \Cpp{} library is provided with the software package, providing an interface to {\Cpp{}} standard-library or ROOT classes, the latter commonly used in high-energy physics.
Both interfaces are described in the following, starting with the text-based interface. The discussion of the text-based interface serves as reference for the description of the \Cpp{} interface.

\subsection{Text-based Interface}
\label{sec:tbui}

The ``convino" executable can be found in the base directory after compiling. It prints usage information and a list of options if the \verb+-h+ option is specified. Other options are:
\begin{itemize}
\item[] \verb+-s+ perform correlation scan
\item[] \verb+-p+ save scan plots as .pdf in addition to a .root file
\item[] \verb+-d+ switch on debug printout
\item[] \verb+--neyman+ uses a Neyman $\chi^2$ instead of the Pearson $\chi^2$
\item[] \verb+--prefix+  defines a prefix for all output files and directories
\end{itemize}
In addition to the options, a text file is passed to the executable. It is referred to as \textit{base file} in the following and is described in Section~\ref{sec:basefile}.
Each measurement comprising one or a set of estimates is described in a \textit{measurement file}. Well documented examples for both types of files are provided in the \texttt{examples} directory and should be consulted alongside this manual.

\subsubsection{Measurement File}

Each measurement file consists of blocks. Each block describes estimates or uncertainties. They are defined by a Hessian, a correlation matrix together with constraints, or a set of orthogonal uncertainties. The latter should be provided in the following format:
\begin{verbatim}
[not fitted]
                  sys_a1  sys_b1  sys_c1    stat   
  estimate_a1     5       6.1     2         6    
  estimate_b1     3       1       4         2
[end not fitted]
\end{verbatim}
The uncertainties \inlineCode{sys\_XX} on the estimates \inlineCode{esti\-mate\_XX} are given in absolute values. The keyword \inlineCode{stat} is reserved for the statistical uncertainty.
The uncertainties and their effect on the estimates in a measurement using a simultaneous nuisance parameter fit technique are described either by a Hessian or a correlation matrix. The Hessian must be written in the following form:
\begin{verbatim}
[hessian]
  sys_a3       1944.6   
  sys_b3      -1349. 1154.4  
  estimate_a3 -0.525 0.5398   3.25e-4    
  estimate_b3 -0.708 0.2706   0       4.89e-4 
[end hessian]
\end{verbatim}
while the correlation matrix has to include additional information about the constraints on the parameters. These constraints are given in units of $1\sigma$ variations for the systematic uncertainties, such that a value of 1 corresponds to no reduction and lower values indicate a constraint from the fit to the data. For estimates, the constraints are given in absolute units and correspond to the total uncertainties. In both cases, they are defined in parentheses, such that the correlation matrix is of the format:
\begin{verbatim}
[correlation matrix]
  sys_a2       (1)     1
  sys_b2       (1)     -0.2         1
  estimate_a2  (10.6)  0.547945 0.1059  1
  estimate_b2  (12.8)  0.147945 0.4305  0 1
[end correlation matrix]
\end{verbatim}
If uncertainties have been described in form of a Hessian or correlation matrix, additional contributions from orthogonal uncertainties can be provided in the \inlineCode{[not fitted]} block. These uncertainties must not have any correlation with the uncertainties defined in the Hessian or the correlation matrix. 
The next block of the measurement file describes the type of each uncertainty.
\begin{verbatim}
[systematics]
    sys_a2 = absolute
    sys_b2 = relative
[end systematics]
\end{verbatim} 
The type can be either \inlineCode{ab\-so\-lute} or \inlineCode{re\-la\-tive}. The default is \inlineCode{ab\-so\-lute} and does not need to be specified explicitly.
The last block defines which of the parameters are estimates, and their nominal values:%
\begin{verbatim}
[estimates]	
    n_estimates = 2
    name_0     = estimate_a2
    value_0    = 780
    name_1     = estimate_b2
    value_1    = 280
[end measurements]
\end{verbatim}
Here, \inlineCode{n\_es\-ti\-mates} gives the number of estimates.

\subsubsection{Base File} 
\label{sec:basefile}
The first block of the base file defines the number of measurement files (\inlineCode{nFiles}) to be considered for the combination and the corresponding file names. An example is given below:
\begin{verbatim}
[input]
    nFiles = 2
    file0   = exampleMeasurement1.txt
    file1   = exampleMeasurement2.txt
[end input]
\end{verbatim}
The files must be in the same directory as the base file.
The second block defines the observables, the estimates should be combined to:
\begin{verbatim}
[observables]
    combined_a = estimate_a1 + estimate_a2
    combined_b = estimate_b1 + estimate_b2
[end observables]
\end{verbatim}
Here, \inlineCode{es\-ti\-mate\_a1} and \inlineCode{es\-ti\-mate\_a2} should be combined to \inlineCode{com\-bined\_a}, and similarly for   \inlineCode{es\-ti\-mate\_b1} and \inlineCode{es\-ti\-mate\_b2}. The number of estimates that should be combined to a single quantity is not limited, as well as the number of combined values. This makes it possible to combine simultaneously e.g. a large amount of bins from differential cross sections from various channels and experiments. However, in this case, the \Cpp{} interface is probably more practical.

The last block describes the correlations that should be assigned using the following syntax.
\begin{verbatim}
[correlations]
    sys_b1 = (0.2) sys_c2
    sys_c1 = (-0.3) sys_d2
[end correlations]
\end{verbatim}
Here, a correlation coefficient of 0.2 is assigned between \inlineCode{s\-ys\_b1} and \inlineCode{s\-ys\_c2} and  -0.3 between \inlineCode{s\-ys\_c1} and \inlineCode{s\-ys\_d2}. 
The correlation assumptions between the parameters can be scanned in an automated way. In this case, the following syntax is used to define the scan ranges:
\begin{verbatim}
[correlations]
    sys_b1 = (0.2 & -0.1 : 0.4) sys_c2
[end correlations]
\end{verbatim}
Here,  \inlineCode{s\-ys\_b1}  has a nominal correlation of 0.2 to  \inlineCode{s\-ys\_c2} . The correlation is scanned from -0.1 to 0.4. 
If several correlation coefficients should be scanned simultaneously, they have to be specified in a single line:
\begin{verbatim}
[correlations]
sys_b1=(0.2&-0.1:0.4)sys_c2+(-0.3&0.2:-0.3)sys_d2
[end correlations]
\end{verbatim}
In this case, the scan range for a single coefficient can start from positive values to negative values to allow accounting for anti-correlations between the parameters that are scanned simultaneously.

Correlation matrices are positive definite by definition, a correlation matrix $C$ with large off-diagonal entries might lose this property if ill-posed assumptions are made, such as:
\begin{equation}
\label{eq:wrongcorr}
C = \begin{pmatrix} 
1 & .99 & 0  \\ 
.99 & 1 & 0.5 \\
0 & 0.5 & 1 \\ 
\end{pmatrix}\text{.}
\end{equation}
In this case, the program exits and it is strongly advised to revise the plausibility of the correlation assumptions.

\begin{figure}[h]
  \begin{center}
\includegraphics[width=\figwidth\textwidth]{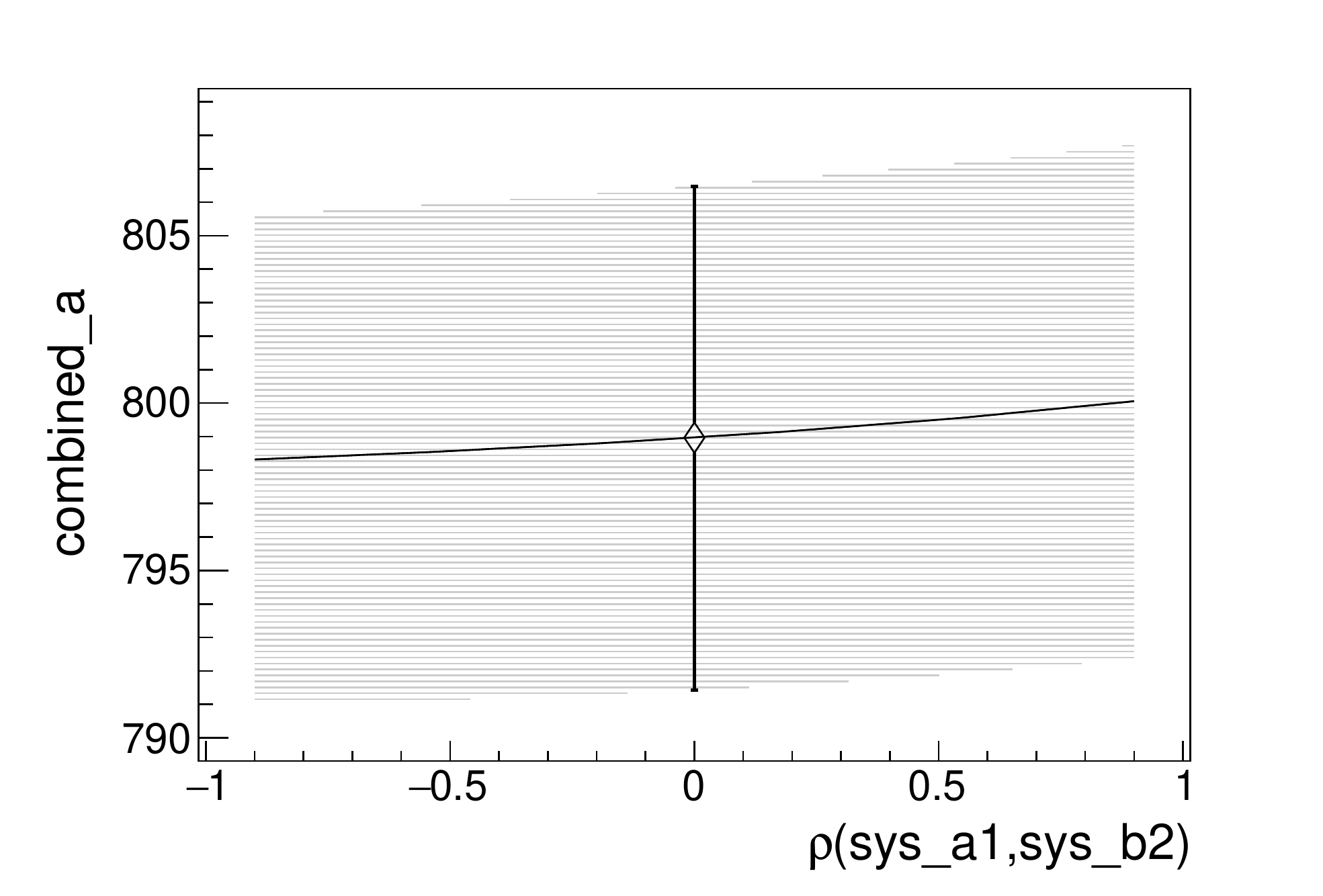}
\end{center}
\caption{Combined value for \inlineCode{com\-bined\_a} 
for a scan of the correlation coefficient for \inlineCode{sys\_a1}
and \inlineCode{sys\_b2}.
The open marker shows the result obtained with the nominal assumption and its uncertainty.The shaded area represents the uncertainty associated to the scanned dependence, indicated by a continuous line. All values are obtained with the example configuration. \label{fig:scan}
} 
\end{figure}

\begin{figure}[h]
  \begin{center}
\includegraphics[width=\figwidth\textwidth]{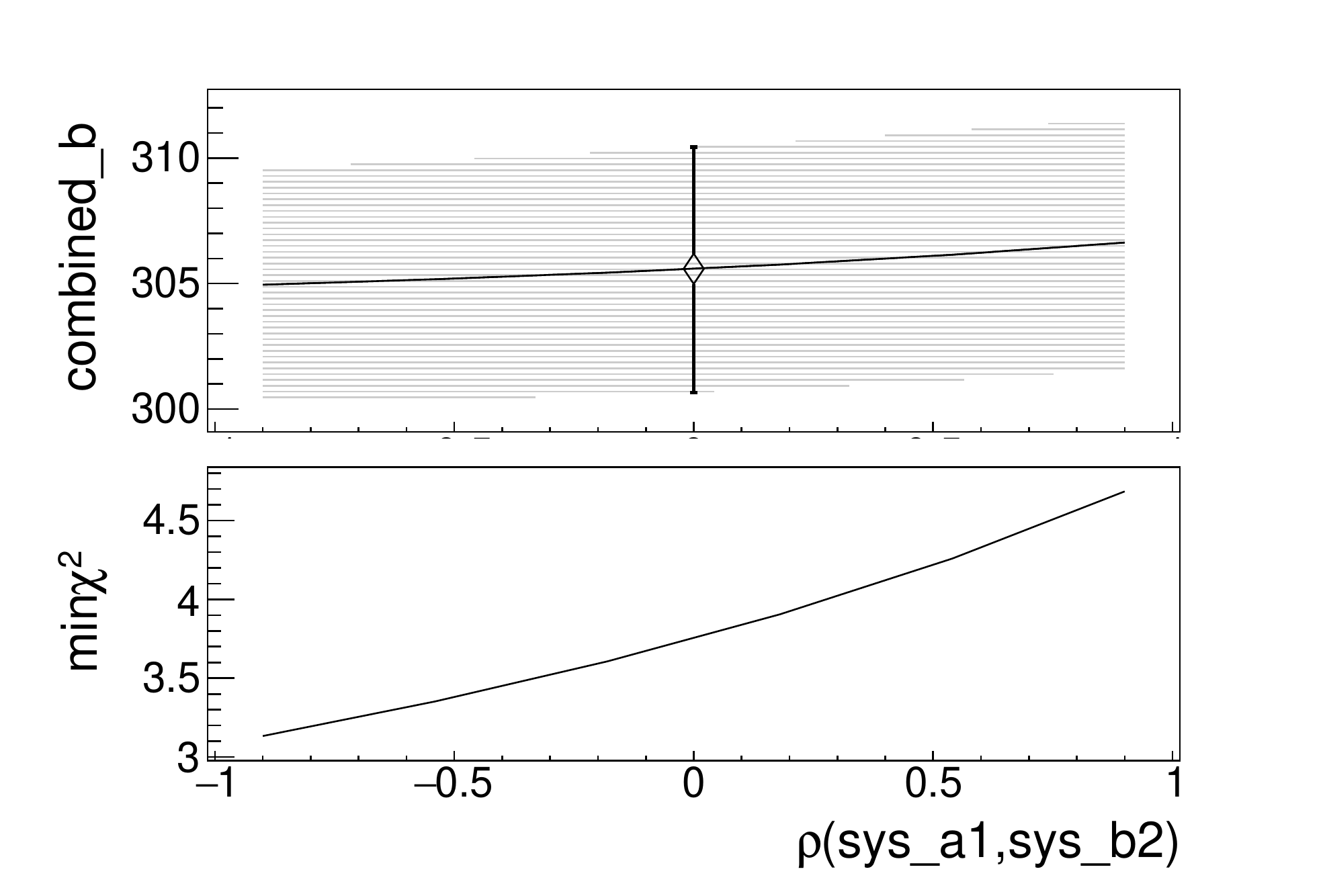}
\end{center}
\caption{Combined values for \inlineCode{com\-bined\_b} 
 (upper panel), and minimum $\chi^2$ (lower panel)  for a scan of the correlation coefficient for \inlineCode{s\-ys\_a1}
and \inlineCode{s\-ys\_b2}.
The open marker shows the result obtained with the nominal assumption and its uncertainty. The shaded area represents the uncertainty associated to the scanned dependence, indicated by a continuous line. All values are obtained with the example configuration. \label{fig:scan2}
} 
\end{figure}

The results of the combination are saved in the output file \path{result.txt}, or \path{<prefix>_result.txt} in case a prefix is specified.
The output file contains the original input correlations, the combination results, the minimum $\chi^2$, and pulls and constraints on all parameters.
The output of the scan, including all correlation matrices, is saved in the file \path{scan_result.txt}. The corresponding figures are saved as \inlineCode{TGraph\-Asymm\-Errors} classes in the file \path{scanPlots.root}. If pdf-file output was enabled, the resulting Figures can be found in the directory \path{scan_results}.
Examples of such Figures obtained with the example configuration are shown in Figure~\ref{fig:scan} and~\ref{fig:scan2}.

\subsection{C++ Interface}

The \Cpp{} interface is optimized for the combination of differential distributions and provides three basic classes which will be described in the following: the class \inlineCode{meas\-ure\-ment}, which is analogous to a measurement file discussed in the previous Section, the class \inlineCode{com\-bi\-ner} to perform the combination, and a class \inlineCode{com\-bi\-na\-tion\-Re\-sult} that collects the output of the combination. The \inlineCode{meas\-ure\-ment} class and the \inlineCode{com\-bi\-na\-tion\-Re\-sult} class provide interfaces to \Cpp{} standard library \inlineCode{std::vec\-tor\-<dou\-ble>} or alternatively to ROOT histograms and graphs.
An example of the usage is provided in \inlineCode{bin/\-differ\-ential\-Ex\-ample.cpp}. Any cpp file that will be placed in the \path{bin} directory will be compiled automatically when running \path{make}.
Alternatively, the compilation of the \programname{} package will create the library \path{libconvino.so} that can be linked against. The header files can be found in the \path{include} directory.

Each class is documented in the corresponding header file. Therefore, the documentation here is limited to the general usage.

\subsubsection{Measurement class}

The measurement class provides the possibility to define a set of estimates, their statistical correlations and systematic uncertainties.
Each object can only contain one set of estimates at once. In case the information is read from a ROOT \texttt{TH1} histogram, each measurement class object can contain only one nominal histogram.

For a measurement with orthogonal uncertainties, the following procedure should be applied:
the nominal values are set using the function \inlineCode{set\-Meas\-ured}. 
Systematic uncertainties can be added in a second step to the measurement object with \inlineCode{add\-Sys\-te\-matics}. 
The type of each uncertainty is defined using the function using \inlineCode{set\-Para\-meter\-Type} after all uncertainties have been added. Here, it is recommended to use the parameter name to identify the correct uncertainty.
In a last step, statistical correlations between the estimates can be set using the method \inlineCode{set\-Es\-ti\-mate\-Cor\-re\-lation}.

If a measurement comprises correlated uncertainties, the corresponding measurement object should be configured using the function \inlineCode{set\-Hes\-sian}, which defines the uncertainties and estimates at once. Additional orthogonal uncertainties can be added in a subsequent step using \inlineCode{add\-Sys\-te\-ma\-tics}.

\subsubsection{Combiner class}

Once the individual \inlineCode{meas\-ure\-ment} objects are defined, they are added to a \inlineCode{com\-bi\-ner} object using the function \inlineCode{add\-Mea\-sure\-ment}.
For the following combination, it is assumed that the entries of each measurement in the same bin or with the same vector index should be combined. It is not possible to combine a number of estimates from one measurement object with a different number of estimates from another. %If this is necessary, it is recommended to create an estimate with a very large statistical uncertainty
The correlation assumptions are defined with \inlineCode{set\-Syst\-Cor\-rel\-ation}. It is advised to use the uncertainties names  as input for unambiguous identification.

The combination is initiated by calling the method \inlineCode{com\-bi\-ne}, which returns a \inlineCode{com\-bi\-na\-tion\-Re\-sult}  class object.

\subsubsection{CombinationResult class}

The \inlineCode{com\-bi\-na\-tion\-Re\-sult} class is a container for all information regarding the inputs to the combination, the correlation matrices, and the combined values as well as the post-combination correlation matrices, pulls and constraints.
If differential distributions are combined, the result can be fed back to a ROOT \inlineCode{TH1} object or a \inlineCode{TGraph\-Asymm\-Errors} using the functions \inlineCode{fill\-TH1} or \inlineCode{fill\-TGraph\-Asymm\-Errors}.

\section{Summary}
\label{sec:summary}

The combination method presented in this document allows combining measurements obtained with simultaneous nuisance parameter fits consistently, taking into account the constraints from the data as well as correlations between systematic uncertainties within each measurement.
In contrast to the optimal case of a direct likelihood combination, based on the product of the individual likelihoods of each measurement, the method does not require the full knowledge of the original data and the fit models. This information would also be required by other commonly used combination methods, however, it is publicly available only in rare cases. It is shown that not accounting for correlations between uncertainties within the same measurement can lead to non-negligible deviations from the combined likelihood approach with respect to the combined value and its uncertainty.

The method described here does not introduce such deviations and relies on the central results and their covariances or Hessians, only, which makes it applicable to a significantly larger variety of combinations.
An extensive validation is performed using {\PM} with varying contributions of statistical and systematic uncertainties, correlation assumptions, binning choices, and {pri-or} statistical correlations. All obtained results and uncertainties are numerically equivalent to a direct likelihood combination. Only for measurements strongly limited by statistical precision, the same known caveats as in other $\chi^2$ or least-squares-based approaches (e.g. the BLUE method) apply. 
In addition, the Convino program is presented. It is developed to perform combinations using the method described here and provides a text-based and a \Cpp{} user interface. The text-based user interface provides an automatic scan of correlation assumptions and creates the corresponding figures for graphical representation.

\section*{Acknowledgements}
I would like to thank O. Behnke for fruitful discussions and comments alongside M. Aldaya Martin, C. Diez Pardos, M. Mulders, and A. Giammanco for editorial suggestions.

\bibliographystyle{ieeetr}
\bibliography{convino}

\end{document}